\documentclass[preprint,12pt]{aastex}

\begin{document}

\title{The Orbit of the Orphan Stream}

\author{
Heidi Jo Newberg\altaffilmark{\ref{RPI}},
Benjamin A. Willett\altaffilmark{\ref{RPI}}, 
Brian Yanny\altaffilmark{\ref{FNAL}}, \&
Yan Xu\altaffilmark{\ref{NAOC}}
}

\altaffiltext{1}{Dept. of Physics, Applied Physics and Astronomy, Rensselaer
Polytechnic Institute Troy, NY 12180, newbeh@rpi.edu\label{RPI}}

\altaffiltext{2}{Fermi National Accelerator Laboratory, P.O. Box 500, Batavia,
IL 60510\label{FNAL}}

\altaffiltext{3}{National Astronomical Observatories of China, 20 Datun Road, Beijing, China\label{NAOC}}

\shortauthors{Newberg, Willett, \& Yanny}

\begin{abstract}

We use recent SEGUE spectroscopy and SDSS and SEGUE imaging data to measure the
sky position, distance, and radial velocities of stars in the tidal debris stream
that is commonly referred to as the ``Orphan Stream."  We fit orbital parameters to the data,
and find a prograde orbit with an apogalacticon, perigalacticon, and eccentricity 
of 90 kpc, 16.4 kpc and $e=0.7$, respectively.  Neither the dwarf galaxy
UMa II nor the Complex A gas cloud have velocities consistent with a kinematic association 
with the Orphan Stream.  It is possible that Segue-1 is associated with the Orphan Stream,
but no other known Galactic clusters or dwarf galaxies in the Milky Way lie along its
orbit.  The detected portion of the stream ranges
from 19 to 47 kpc from the Sun and is an indicator of the mass interior to these distances.
There is a marked increase in the density of Orphan Stream stars near 
$(l,b)=(253^\circ,49^\circ)$, which could indicate the presence of the progenitor at the 
edge of the SDSS data.  If this is the progenitor, then the detected portion of the 
Orphan Stream is a leading tidal tail.  We find blue horizontal branch (BHB) stars and 
F turnoff stars associated with the Orphan Stream.  The turnoff color is $(g-r)_0=0.22$.  
The BHB stars have a low metallicity of [Fe/H]$_{\rm WBG}=-2.1$.  The orbit is best fit 
to a halo potential with a halo plus disk mass of about $2.6 \times 10^{11} M_\odot$, 
integrated to 60 kpc from the Galactic center.  Our fits are done to orbits rather than
full $N$-body simulations; we show that if $N$-body simulations are used, the inferred
mass of the galaxy would be slightly smaller.  Our best fit is found with a logarithmic 
halo speed of $v_{\rm halo}=73\pm 24~ \rm km~s^{-1}$, a disk+bulge mass of
$M(R< 60 {\rm ~kpc}) = 1.3 \times 10^{11} M_\odot$, and a halo mass of
$M(R< 60 {\rm ~kpc}) = 1.4 \times 10^{11} M_\odot$.  However, we can find similar fits
to the data that use an NFW halo profile, or that have smaller disk masses and 
correspondingly larger halo masses.  Distinguishing between different classes of models
requires data over a larger range of distances.  The Orphan Stream is projected to extend 
to 90 kpc from the Galactic center, and measurements of these distant parts of the stream 
would be a powerful probe of the mass of the Milky Way.

\end{abstract}


\section{Introduction}

The discovery that the Milky Way's stellar halo is lumpy \citep{netal02} has sparked a flurry of
activity in identifying the Milky Way's stellar substructure.  Dwarf galaxies and globular clusters
undergo tidal disruption, creating streams of stars in the halo.  New, low surface brightness dwarf
galaxies and globular clusters have been identified 
\citep{2005AJ....129.2692W,2007ApJ...669..337K,2007ApJ...654..897B}, and several large 
structures have been discovered
whose nature is unknown or controversial \citep{ynetal03, triand, belok07, juric08}.  
Tidal debris in the Milky Way's stellar halo teaches us
about the accretion history of the Milky Way, and it has been suggested that tidal debris may be the
key to measuring the distribution of dark matter in the Galaxy \citep{cetal99,md99,h04, 
jlm05,eb2009,MajewskiSIM2009,oetal09}.

One of the newly discovered substructures is the ``Orphan Stream," which was independently discovered
by \citet{getal06} and \citet{betal06a}.  \citet{getal06} was the first to publish a full discovery
paper, showing that the stream was at least $60^\circ$ long and $2^\circ$ wide, and likely the remains of a small dwarf
galaxy that has been completely disrupted.  In Grillmair's paper, the stream was characterized as 
about 21 kpc distant from the Sun along the whole length of the stream.  \citet{betal07} published
an independent discovery paper naming the Orphan Stream; they found a similar length, width, and 
likely origin.  However, Belokurov et al. found a strong distance gradient, from 20 kpc at one end 
of the stream to 32 kpc from the Sun at the other end.  They also published ``suggestive" radial velocities 
from sparse samples of Sloan Digital Sky Survey (SDSS) data in two fields, ranging from $V_R =$ -40 km s$^{-1}$ at the close end of 
the stream to $\sim 100$ km s$^{-1}$ at the distant end of the stream.  They noted that the dwarf 
galaxy Ursa Major II, the HI clouds of Complex A, and a number of anomalous globular clusters
(including Ruprect 106 and Palomar I) lie near the same great circle as the Orphan Stream stars.

Following the discovery papers, three further studies,
\citet{fetal07}, \citet{jl07}, and \citet{setal09}, attempted to fit an orbit the the Orphan Stream
and to determine whether it was connected to the assortment of objects on the same great circle. 
\citet{fetal07} fit two tidal disruption models to the available Orphan Stream data, with the
assumption that the UMa II dwarf galaxy is the progenitor of the stream.  They successfully fit
the position and distance data, and the velocity for the more distant part of the stream.  They
are only able to fit the near velocity, at $V_R=-40$ km s$^{-1}$, by assigning part of the
Orphan Stream to the leading tail and part of the Orphan Stream to the trailing tail.  They conclude
that UMa II is a likely progenitor of the Orphan Stream, that some of the young halo globular clusters
may be associated with the Orphan Stream, that Complex A could be matched only if it was a whole orbit
ahead of the Orphan Stream, and that the picture made the most sense if all of these objects came from
an original object that was a larger dwarf galaxy.
Shortly after the Fellhauer et al. paper, \citet{jl07} explored the possibility that Complex A and 
the Orphan Stream were related to each other.  They used the sky positions of the Orphan Stream
and Complex A and the radial velocities of Complex A, assumed the two objects were along the same
orbit, and fit the values for the distances to the two objects and the radial velocities along the
Orphan Stream.  In order to make the orbit fit, the distance to Complex A is 4 kpc, the distance to
the Orphan Stream is assumed to be 9 kpc, and the heliocentric radial velocity of the Orphan Stream ranges from 
-155 km s$^{-1}$ to -30 km s$^{-1}$.  These values are in conflict with the measurements of previous
authors and this paper, so it seems unlikely that the Orphan Stream and Complex A are on the same
orbit, and in the same revolution of orbital phase.  More recently, \citet{setal09} successfully fit 
an orbit to the distances and velocities measured in the \citet{betal07} paper.  They concluded that
the orbit of the Orphan Stream does not intersect UMa II or Complex A, and that the data is
consistent with ``a single wrap of the trailing arm of a fully disrupted satellite."

The fits to the Orphan Stream orbit depend on having reliable measurements of the star positions, distances, and (especially) velocities along the Orphan Stream.  In this paper, we extract the best distances and velocities of
Orphan Stream stars in the SDSS Data Release 7 (DR7; Abazajian et al. 2009), making extensive use of the newly released Sloan Extension for Galactic Understanding and Exploration (SEGUE) subsample of
stellar velocities \citep{yetal09a}.  We show that the suggestive \citet{betal07} velocities, which were used in the orbit calculations of
later authors, were not representative of true Orphan Stream members.  

We fit a new orbit to the Orphan Stream data that is consistent with our new measurements, and
find a best fit Milky Way mass of $M(R<60 {\rm~kpc}) = 2.65 \times 10^{11}M_\odot.$
If we extrapolate the integrated mass out to a radius 240 kpc (approximately the virial radius), 
assuming a log potential normalized to $v_{\rm halo} = 73 \rm ~km~s^{-1}$,
we find $M(R < 240 \rm ~kpc) \sim 6.9\times 10^{11} M_\odot$.
A recent history of mass estimates for the Milky Way include
\citet{1999MNRAS.310..645W}, who use the orbits of satellite galaxies
and globular clusters and find $M(R<50\rm ~kpc) = 5.4^{+0.2}_{-3.6} \times 10^{11} M_\odot$ and $M_{\rm virial} = 1.9^{+3.6}_{-1.7}\times 10^{12} M_\odot$.
\citet{2002ApJ...573..597K}, on theoretical formation grounds predict
a virial mass for the Milky Way $M_{\rm virial} \sim 1\times 10^{12} M_\odot$, almost
twice as high as our measurement.
Other estimates are from \citet{2005MNRAS.364..433B} with 
$M_{\rm virial} = 0.8^{+1.2}_{-0.2}\times 10^{12} M_\odot$ from halo tracers and 
\citet{2008MNRAS.384.1459L} with $M_{\rm virial} = 2.43\times 10^{12} M_\odot$ from theory and
relative motions of the Milky Way and Andromeda.
\citet{2006MNRAS.365..747A} used a careful combination of
predicted velocity dispersion theory and observations from \citet{2005MNRAS.364..433B} to
arrive at $M_{\rm virial} \sim 6.4\times 10^{11} M_\odot$, consistent with the value we 
propose here.  Furthermore, \citet{2007MNRAS.379.1475S} propose a relation, based on 
analysis of large N-body simulations of satellites and their hosts, between the dispersions 
of the satellite velocities and the virial velocity of each host. Applied to the Milky Way, 
they derive a virial mass of $7 \times 10^{11} M_\odot$.  Our mass estimate is on the low
end, but within the range of recent measurements of the Milky Way's mass.

\section{Angular position of the Orphan Stream}

We begin with an analysis of the positions of stream stars in the sky,
as the sky positions of the Orphan Stream are the least controversial measurements.
We select stars from SDSS and SEGUE photometric data.

The SDSS Legacy survey imaged one quarter of the 
sky in $ugriz$ filters uniformly (photometric color 
errors of $<2$\% to $r\sim 21$) to faint depth ($g \sim 23$).  
Additional imaging was carried out as part of the SEGUE survey.
Simple color cuts allow one to isolate large populations and 
construct density maps of certain selected spectral types. 
Magnitude-position space maps of blue horizontal branch (BHB), F turnoff and 
red giant branch (RGB) stars, being more or less accurate 
standard candles, are a tool 
for isolating substructure in stars in our Milky Way's halo \citep{netal02}.  
We deredden all halo star candidate photometry (indicated with a `0' subscript on magnitudes and colors), which are generally beyond the foreground 
dust screen, using the standard maps of \citet{sfd98}.

We selected from the DR7 database of STARs, all objects with $100^\circ<\alpha<175^\circ$ and
$-25^\circ<\delta<70^\circ$.  Later in the paper, we will show that the stars of the Orphan Stream
are at distances between 15 and 50 kpc from the Sun, and have a turnoff color near $(g-r)_0=0.22$.  Therefore,
to select from this dataset all of the stars that could be Orphan Stream turnoff stars, we selected
only those stars with $0.12<(g-r)_0<0.26$, $(u-g)_0>0.4$, and $20.0<g_0<22.5$.  Redder stars were not
selected because there is a larger background of field disk and halo stars on the red side of the turnoff than on the blue side.
A greyscale plot of
the sky densities of these stars, presented in Galactic coordinates, is shown in Figure 1.  The most
prominent excess of F turnoff stars is from the leading tidal tail of the Sagittarius dwarf
spheroidal galaxy, and a fainter stream, almost parallel to the first that may be a bifurcation
of the Sagittarius dwarf tidal stream \citep{betal06a}.  The narrower, more horizontal, tidal debris
stream is the Orphan Stream.  Note that the surface density of stars in the Orphan Stream increases just
before the left edge of the SDSS data.  Since this
portion of the stream is closer to the Galactic center (see below), we do not expect the density to increase as it
does at apogalacticon where the velocities are lower.  The increase in density may indicate the
progenitor is at the edge, near $l=253^\circ$, or just off the edge of the data 
on the left side, towards higher Galactic longitudes.

We trace the position of the Orphan Stream across the sky, and tabulate the Galactic latitude
every ten degrees in Galactic longitude along the stream in Table 1.  The center of the stream was
determined to the nearest half degree pixel, so the error in latitude is estimated at about 0.7 degrees.
We then calculated the Euler angles required to transform from Galactic coordinates to a Sun-centered
coordinate system where the portion of the Orphan Stream that we detected is on the equator.
Using a notation similar to that used by \citet{mswo03} to describe the Sagittarius dwarf orbital
plane, we define $B_{\rm Orphan}$ and $\Lambda_{\rm Orphan}$ by the rotation:
\[ \left( \begin{array}{c}
\cos{B_{\rm Orphan}} \cos{\Lambda_{\rm Orphan}} \\
\cos{B_{\rm Orphan}} \sin{\Lambda_{\rm Orphan}} \\
\sin{B_{\rm Orphan}} \end{array} \right) =
{\mathcal M}
\left( \begin{array}{c}
\cos{b} \cos{l} \\
\cos{b} \sin{l} \\
\sin{b} \end{array} \right), \]
where
\[{\mathcal M} = \left( \begin{array}{ccc}
\cos{\psi}\cos{\phi}-\cos{\theta}\sin{\phi}\sin{\psi} & \cos{\psi}\sin{\phi}+\cos{\theta}\cos{\phi}\sin{\psi} & \sin{\psi}\sin{\theta} \\
-\sin{\psi}\cos{\phi}-\cos{\theta}\sin{\phi}\cos{\psi} & -\sin{\psi}\sin{\phi}+\cos{\theta}\cos{\phi}\cos{\psi} & \cos{\psi}\sin{\theta} \\
\sin{\theta}\sin{\phi} & -\sin{\theta}\cos{\phi} & \cos{\theta} \end{array} \right). \]
The values of $(\phi, \theta, \psi) = (128.79^\circ, 54.39^\circ, 90.70^\circ)$ were determined
by using gradient descent to find the fit that minimized the sum of the squares of the angular
distances between the equator at $B_{\rm Orphan}=0^\circ$ and the first nine stream positions in 
Table 1 (the last point was added after further study discussed in \S 4).  The
zeropoint of $\Lambda_{\rm Orphan}$ was set at $l=220^\circ$, near where the leading tidal tail of
the Sagittarius dwarf crosses the Orphan Stream.  In the lower panel of Figure 1, we show that the line $B_{\rm Orphan}=0^\circ$ 
passes through the center of the Orphan Stream in the region where we detect it.
The $(\Lambda_{\rm Orphan}, B_{\rm Orphan})$ values for points along the Orphan Stream are also
tabulated in Table 1.
Note that at low Galactic longitude, the stream departs slightly from $B_{\rm Orphan}=0^\circ$.
It is not possible to fit an orbit over the whole sky with a simple rotation of coordinates, even if the
Galactic potential is spherical, since we view the sky from a Sun-centered, rather than Galaxy-centered,
viewpoint.  In general, halo orbits do not follow great circles over their entire length.

\section{Distance to the Orphan Stream}

Although the large number of F turnoff stars make them the best tracers of the position of the
stream in the sky, their large range of absolute magnitudes at a given distance make them less useful as distance
indicators.  Instead, we start by photometrically selecting blue horizontal branch (BHB) stars 
along the stream.  We color-select A stars from all of the STARs in SDSS DR7 that are within one degree of 
the center of the stream ($-1^\circ<B_{\rm Orphan}<1^\circ$), using the color range 
$-0.4<(g-r)_0<0.0$, $0.8<(u-g)_0<1.6$, and the magnitude range $15<g_0<21.5$.  In 
\citet{ynetal00} we showed that low and high surface gravity A stars could be roughly separated 
from each other using their $ugr$ colors, as shown in Figure 10 of that paper; low surface gravity 
BHB stars are redder in $(u-g)_0$ at a given $(g-r)_0$.   We perform this separation on the color-selected A stars.

In Figure 2 we show $g_0$ vs. $\Lambda_{\rm Orphan}$ 
for the 593 stars with colors of BHB stars and within one degree of the path of the Orphan 
Stream in the $\Lambda_{\rm Orphan}$ range showed.  
We argue below that an excess of BHBs exists along the line shown, and
have added spectrosopic Orphan Stream candidates to the figure here as
open circles.
Above the locus of Orphan Stream BHBs stretching from $g_0=17$ at 
$\Lambda_{\rm Orphan}=20^\circ$ to $g_0=19$ at $\Lambda_{\rm Orphan}=-25^\circ$, there is a set of what appears to be blue straggler stars that, 
due to imperfect $ugr$ color separation, leaked into our selection box 
two magnitudes fainter \citep{ynetal00}.
Our distances to positions along the Orphan Stream agree well with those of
\citet{betal07} where the data directly overlap.

We now refine the distance to the stream using SDSS/SEGUE spectroscopy of BHB stars.  
Because we can rely on the spectroscopy to separate low from high
surface gravity (measured by the tabulated DR7 parameter $\log \rm g$) A stars, we can detect horizontal branch stars that are redder than $(g-r)_0=0$,
where the photometric separation becomes ineffective.  
We selected
all spectra from SDSS DR7 that were within $1^\circ$ of the Orphan Stream 
($-1^\circ<B_{\rm Orphan}<1^\circ$), in the range $-30^\circ <\Lambda_{\rm Orphan} < 30^\circ$ where the the stream is
visible in F turnoff stars, in the color range $-0.4<(g-r)_0<0.3$ and $0.8<(u-g)_0<1.6$,
and with $0.1< \log \rm g<3.5$.  
We further required that the sppParams table (SSPP, see Section 6) parameter 
``elodierverr" be greater than zero, indicating that the pipeline successfully fit a stellar 
template to the spectrum.  This removes QSOs and galaxies from the sample.  Starting with this 
sample, we made a series of plots of $v_{gsr}$ vs. $\Lambda_{\rm Orphan}$ and $g_0$ vs. 
$\Lambda_{\rm Orphan}$, and removed stars that were not clustered in both distance and 
velocity.  In other words, if stars in an apparent velocity clump were not also clustered in 
apparent magnitude, then that group was removed from the sample.  We also discovered that by 
selecting stars with low metallicity, [Fe/H]$_{\rm WBG}<-1.6$ \citep{wbg99}, we lost fewer stream
stars than background.  The BHB stars that were left after this iterative selection process are plotted
in Figure 2 as open circles.

We measure the apparent magnitude of BHB stream stars in five places where there is enough data
in Figure 2 to be fairly confident of the stream location; these magnitudes are tabulated in Table
1, along with estimates of the error.  These error estimates are rough estimates based on how
much we believed we could vary the magnitude and still stay consistent with the datapoints
on the stream.  We then estimate the function $g_0(\Lambda_{\rm Orphan})$
for Orphan Stream BHB stars by fitting a parabola to the four datapoints with lower errors.  The best fit is
\[g_0 = 0.00022 \Lambda_{\rm Orphan}^2 - 0.034 \Lambda_{\rm Orphan} + 17.75. \]

\section{Color-Magnitude Diagram of the Orphan Stream}

We can now use the relationship between the $g_0$ magnitude of the horizontal branch and 
$\Lambda_{\rm Orphan}$ to adjust the apparent magnitudes of stars along the Orphan Stream so 
that the horizontal branch is at a constant magnitude across its entire extent.  We do this
by calculating $g_{corr}=g_0 - 0.00022 \Lambda_{\rm Orphan}^2 + 0.034 \Lambda_{\rm Orphan}$.  Note that 
the position of the horizontal branch at $\Lambda_{\rm Orphan} = 0^\circ $ remains unchanged.  We select all 
of the STARs from SDSS DR7 that are within $2^\circ$ of the Orphan Stream 
($-2^\circ<B_{\rm Orphan}<2^\circ$) and in the portion of the stream where it is
visible in Figure 1 and away from the Sagittarius stream ($-26^\circ<\Lambda_{\rm Orphan}<-7^\circ$ or 
$10^\circ<\Lambda_{\rm Orphan}<40^\circ$).  Background stars were selected from the same
regions of $\Lambda_{\rm Orphan}$, but with $2^\circ < |B_{\rm Orphan}|<4^\circ$.

In Figure 3, we show a background subtracted Hess diagram of stars along the Orphan Stream.
In order to better see the horizontal branch stars, we include an insert of 
the BHB section of the `on stream' ($|B_{corr,Orphan}| < 1$) color magnitude diagram 
(without background subtraction) in the regions of the diagram in which the color 
sub-selection is valid.  
Here, the Orphan BHBs are clearly seen at $g_{\rm corr} = 17.75$ as an overdensity. 
There is also a large overdensity of F turnoff stars at $(g_0, (g-r)_0)=(21.2, 0.22)$.

We are now in a position to `correct' the $g_0$ magnitudes of the stars in Figure 1 so
that turnoff stars along the entire length of the Orphan Stream will be at the same
magnitude.  This allows us to make a narrower magnitude cut and see the stream over
a larger range of distances.  In Figure 4 we show the $(l,b)$ distribution of the F 
turnoff stars with $0.12<(g-r)_0<0.26$ and $20.7<g_{corr}<21.7$.  This allows us to 
see the Orphan Stream all the way to $l=170^\circ$.  The position of the stream at 
$l=171^\circ$ was added to Table 1 from analysis of Figure 4.  The 
upper panel of Figure 4 shows that the Orphan Stream is several degrees above
$B_{\rm Orphan}=0^\circ$ at low Galactic longitude.  

This diagram
suggests that for $l<200^\circ$, we should search for Orphan Stream
stars over a higher range of $B_{\rm Orphan}$.  
In order to follow the Orphan Stream at lower Galactic latitudes, it is thus necessary to correct 
$B_{\rm Orphan}(\Lambda_{\rm Orphan})$, since the stream is not at $B_{\rm Orphan}=0^\circ$ in this
region.  We fit a parabola to the to the three points at $l=175^\circ, 185^\circ, \rm ~and~ 195^\circ$
in Table 1, to find $B_{\rm Orphan} = -0.00628 \Lambda_{\rm Orphan}^2 - 0.42 \Lambda_{\rm Orphan}
-5.00$ for $-35^\circ < \Lambda_{\rm Orphan} < -15^\circ$.  We define a new variable, 
$B_{\rm corr}$, for which the Orphan Stream is at $B_{\rm corr}=0^\circ$ along our entire data set.
$B_{\rm corr}=B_{\rm Orphan}$ for $\Lambda_{\rm Orphan} \ge -15^\circ$, and 
$B_{\rm corr}=B_{\rm Orphan}+0.00628 \Lambda_{\rm Orphan}^2 + 0.42 \Lambda_{\rm Orphan} + 5.00$
for $\Lambda_{\rm Orphan}<-15^\circ$.

In order to search for the Orphan progenitor in the gap between $255^\circ < l < 268^\circ$ where 
there is no SDSS data available, we searched the online SuperCOSMOS data set 
\citep{2001MNRAS.326.1279H} for faint turnoff stars in the gap.
We selected stars with $ 0.5 < B_J-R_2 < 1.0, 20 < B_J < 21$ and added them 
to Figure 4 with approximately the same
scaling and the same orientation as the SDSS/SEGUE imaging.  A faint,narrow
trail, likely the extension of the Orphan stream tidal tail, is visible 
at $255^\circ < l < 268, 38^\circ < b < 48^\circ$.
The SuperCOSMOS data suffers from poor photometric calibration across and between
measured photographic plates, which makes it impossible to trace the Orphan Stream
in the region in which we detect it in the SDSS data.  This makes it difficult for us to calibrate
the density of stars in the SuperCOSMOS detection with the density observed in the
SDSS data.  We are unable to either confirm or rule out the overdensity at $(l,b)=(253^\circ,49^\circ)$.

\section{Density of F turnoff stars along the Orphan Stream}

In Figure 5, we show the number of F star counts above background, as a function of position along
the Orphan Stream.  The stream stars were selected to be within
$-2^\circ<B_{\rm corr}<2^\circ$.  The background stars were selected from the region of the
sky that is between $2^\circ$ and $4^\circ$ in angular distance from the stream (including
stars on both sides of the stream).  The binning of Figure 5 was chosen so that `edge-of-bin' effects do not dominate the relative number counts in the `outrigger' SEGUE stripe at $l=270^\circ $, ($\Lambda_{\rm Orphan} = +35^\circ$).

We note from this density plot that the stellar density is much higher near 
$\Lambda_{\rm Orphan}=22^\circ$ than it is along the rest of the observed tidal tail.  This is
very interesting because this portion of the stream is also the closest to the Sun and the
Galactic center.  It is not expected that the stellar density will rise sharply along the
tidal stream, especially closer to the Galactic center, except near the progenitor.  Stream 
stars generally pile up at apogalacticon, where they are moving more slowly.

Therefore, we interpret this as evidence that the progenitor dwarf galaxy is at or near the edge of the SDSS footprint, at $(l,b) = (255^\circ,48^\circ)$.
If this is correct, and the Orphan progenitor is located in the region $255^\circ < l < 268 ^\circ$, then the Orphan Stream stars intersected on ``outrigger'' SEGUE 
stripe 1540 \citep{yetal09a}, 
near $(l,b) = (270^\circ,39^\circ)$ are part of the 
trailing (rather than leading) tail (see below for tail orientation).

The density falls sharply for $\Lambda_{\rm Orphan}<-25^\circ$.  This could be due to a combination
of three factors: (1) the stream density may actually drop, (2) the distance to the 
stream could be extrapolated
incorrectly in this angular range, causing us to miss stream stars in the data outside of our $20.7 < g_{\rm corr} < 21.7$ cut, or (3) the falling completeness of the F turnoff stars in the SDSS DR7 database at $g_0 > 22.5$ reduces the
contrast of the stream against the background.  
Therefore, we can say little with certainty about the stream 
density for $\Lambda_{\rm Orphan}<-25^\circ$.

\section{Velocities of Orphan Stream stars}

Now that we have a measure of the spatial position of the Orphan Stream, we can improve our
selection of spectroscopically observed stars in the stream.  
While initial detection of halo substructure is usually best done with 
deep, wide-field, filled-sky multi-color imaging, determination of the orbit
of a given structure depends on kinematics, either proper motions or
radial velocities (or both), for identified structure members.  Sufficiently 
accurate proper 
motions are currently difficult to obtain for halo objects 
more distant than about 15 kpc from the Sun (unless hundreds or thousands 
of individual proper motions can be averaged), and thus for 
outer halo structures, we rely primarily on radial velocities from
spectroscopy to constrain structure orbits.  

The SEGUE survey contributed 
velocities for 240,000 stars, including several thousand targeted 
spectra of halo stream candidates with velocity accuracies of 5-10 $\rm km~s^{-1}$, sufficient to isolate stream members from the broad background halo and constrain
structure velocity dispersions.  Spectra are also highly useful in
determining structure membership for individual objects by providing estimates of stellar
metallicity and luminosity class (surface gravity).   Diffuse halo stream tidal tails are often overwhelmed in number by unrelated halo stars, but by 
using the SEGUE SSPP pipeline \citep{letal08a,letal08b,apetal08} estimates of
stellar atmospheric parameters, individual stream candidates may be separated
from the background.  

We select A stars from all stars
with spectra in SDSS DR7 using the color selection $-0.4<(g-r)_0<0.3$ and $0.8<(u-g)_0<1.6$.
Throughout this paper, all stellar spectra are selected from the database with 
elodierverr$>0$ to remove galaxies and QSOs from the sample.  We
limit the spectra to BHB stars by selecting those with average surface gravity measurement from 
the SSPP in the range $0.1<\log \rm g<3.5$.  We then select spectra within two degrees of the Orphan
stream.  For $\Lambda_{\rm Orphan}<-15^\circ$, we select stars with 
$-2^\circ < B_{\rm corr} < 2^\circ$.  And
finally, we select BHB stars that have apparent magnitudes consistent with the distance to the
Orphan Stream ($17.4 < g_0 - 0.00022 \Lambda_{\rm Orphan}^2 + 0.034 \Lambda_{\rm Orphan} < 18.0$).

The filled circles in Figure 6 show the line-of-sight, Galactic standard of rest velocities  ($v_{\rm gsr} = \rm RV + 10.1\cos(b)\cos(l)+224\cos(b)\sin(l)+6.7\sin(b)$ km s$^{-1}$) for the spectroscopically observed BHB stars with sky positions and distances consistent with
membership in the Orphan Stream.  Stars with [Fe/H]$_{\rm WBG} < -1.6$ are circled.  The Orphan
Stream stars are clearly separated from the background, at velocities near $v_{\rm gsr}\sim +110$ km s$^{-1}$.  Almost all of the BHB stars that have Orphan Stream velocities also have
[Fe/H]$_{\rm WBG} < -1.6$.  We have highlighted in green the low metallicity stars with
$-0.0445 \Lambda_{\rm Orphan}^2 - 0.935 \Lambda_{\rm Orphan} + 95$ km s$^{-1} < v_{\rm gsr} < 
-0.0445 \Lambda_{\rm Orphan}^2 - 0.935 \Lambda_{\rm Orphan} + 165$ km s$^{-1}$.  These are 
the stars that
we used to determine the velocity of the stream as a function of $\Lambda_{\rm Orphan}$.  
For comparison, the preliminary velocities from \citet{betal07} are marked with magenta squares.

The stars in Figure 6 are binned by $\Lambda_{\rm Orphan}$ in $10^\circ$ bins starting 
at $\Lambda_{\rm Orphan} = -35^\circ$.  Seven bins contain Orphan Stream stars, as shown
in Figure 6.
In each bin we measure the mean $\Lambda_{\rm Orphan}$,
the mean $g_0$, error in the mean ($\delta g_0$),
the mean $v_{\rm gsr}$, error in the mean ($\delta v_{\rm gsr}$), and standard deviation
of the velocity distribution ($\sigma_v$).  These values, and the number of stars in 
each bin ($N$), are tabulated in Table 2.

Because we now have better information on the path of the stream, it is worth revisiting the
magnitude measurements that we originally determined from the data in Figure 2.  In Figure 7 we show
$g_0$ vs. $\Lambda_{\rm Orphan}$ for
the photometrically selected BHB stars (selected the same way as those in Figure 2) that are now
within $2^\circ$ of the Orphan Stream.
We also show the Orphan Stream BHBs with spectra, as selected from Figure 6.
The Orphan Stream can be traced about ten degrees further across the sky using the new data
selection than it could be using Figure 2.  For each value of $\Lambda_{\rm Orphan}$ in Table 2, we use Figure 7 to determine
a value of $g_0$, and estimate the error in this determination.

For each value of $\Lambda_{\rm Orphan}$, we determine the Galactic longitude for 
$B_{\rm corr}=0^\circ$.  Then we use the data in Figure 4 to determine the Galactic latitude
at that longitude.  These numbers are also tabulated in Table 2.  Note that although the basic
data is similar, the determinations of position and distance were made independently, and may
not be identical to those of Table 1, even for similar values of Galactic longitude.

Table 2 shows that the typical velocity dispersion in the stream is 10 km s$^{-1}$.  The
velocity dispersion of the last point, with only 3 stars, at $l=271^\circ$, is unphysically low, since it is
smaller than the velocity errors, but it is the number that was measured using the
technique described.  The position of the stream here was quite difficult to pick out from the 
photometry, due to the gap in the data.  The unexpected velocity dispersion, for stars that
may not be at the same location as the photometric overdensity, leads to a large
uncertainty in the validity of the detection here.  Because the detection is uncertain,
and it is not known if the stars at this position are in the same tail as the
rest of the data, this data
point was excluded from orbit fits for the Orphan Stream (see \S 10).

\section{Orphan Stream spectra}

Now that we know the velocity of the stream as a function of sky position, we can look for stars
other than BHB stars that are part of the Orphan Stream.  We start by selecting all stellar spectra
that are now within $2^\circ$ of the Orphan Stream, following the method outlined in the first
paragraph of \S 6, and are within the portion of the stream that we have observed,
$-35^\circ < \Lambda_{\rm Orphan} < 40^\circ$.  We eliminate sky fibers by choosing only spectra
with $g_0<22$.  We choose a somewhat narrower range of velocities than in our previous velocity
cut, to reduce the number of background stars in the sample.  We select stars with
$-0.0445 \Lambda_{\rm Orphan}^2 - 0.935 \Lambda_{\rm Orphan} + 112.5 < v_{\rm gsr} < 
-0.0445 \Lambda_{\rm Orphan}^2 - 0.935 \Lambda_{\rm Orphan} + 147.5$.

Brighter stars that are consistent with Orphan Stream membership are expected to be giant stars
with low surface gravity, while fainter stars should be higher surface gravity main sequence stars.
To determine which stars are stream star candidates, we calculated 
$g_{\rm corr}$.  This moves all of the stream stars to the distance modulus of the
Orphan Stream at $\Lambda_{\rm Orphan}=0^\circ$.  Then, all of the stars with
$g_{\rm corr} < 19+ 4(g-r)_0$ should be giants, so stars outside of the range
$0.1 < \log \rm g < 3.5$ were discarded.  We restricted the surface gravities of the
fainter stars ($g_{corr}>19+4(g-r)_0)$) to high surface gravities ($\log \rm g>3.0$).  This 
was necessary to reduce the number
of stars that in the blue straggler region of the diagram, but unfortunately also removed all of
the candidate turnoff stars, which were faint enough that they did not
have measured surface gravities.

Our final data selection was for low metallicity stars.  Since these stars are not all BHB stars, we
used the adopted metallicity measurement (feha) in the SDSS database, rather than the [Fe/H]$_{\rm WBG}$
measurement that we used in our previous selection.  We limited our sample to stars with
[Fe/H]$_{\rm adopted}<-1.6$.  The final dataset includes 87 stars, which are shown as magenta crosses in Figure 8.  Figure 8 is a reproduction of Figure 3 with overlays described below.

In order to understand the background spectral distribution, we also select a set of stars with
spectra that were near, but not on, the Orphan Stream.  The on-stream stars were selected to be
within two degrees of the sky position of the stream, with $-36^\circ<\Lambda_{\rm Orphan}<40^\circ$.
The off-stream stars were selected with a distance of two to five degrees from the stream.  The
sky area of the off-stream stars is 50\% larger, but because the SEGUE spectral coverage is not uniform on the sky (there are several SEGUE pointings directly
on the stream), it contains a similar number of SDSS/SEGUE spectra.  Using these sky areas, we selected 
8385 stars in the on-stream sample, and 8726 stars in the off-stream sample.  The off-stream sample was
then culled using identical velocity, surface gravity, and metallicity criteria.  The resulting 
``background" spectra are overlaid in Figure 8 as black circles.

Some of the spectroscopically selected candidate Orphan Stream members show a reasonable correspondence
to the positions we would expect from CMDs of globular clusters.  We see the BHB stars at
$g_{\rm corr} \sim 17.75$.  There are 
candidate blue straggler stars extending from the turnoff at $g_{\rm corr} \sim 21$ towards
brighter magnitudes and bluer colors.  There is a nearly vertical red giant branch at
$(g-r)_0=0.5$.  There are many fewer comparison background spectra left after selecting for Orphan Stream candidate parameters;
these show no horizontal branch, many fewer blue straggler candidates, and the stars near the giant
branch do not show the expected trend of redder stars at brighter magnitudes.

There are two selection effects in this diagram that should be clearly understood.  First, the
surface gravity selection is different for stars below a line that runs from 
$[g_0, (g-r)_0]=(17, -0.5)$ to $(g_0, (g-r)_0)=(21, 0.5)$.  This selection could visually
enhance the appearance of the blue straggler separation.  
The second selection effect is that neither SDSS nor SEGUE spectroscopic selection evenly
samples stars in color or magnitude.  In particular, a very large number of SEGUE spectra were obtained
for G stars with $0.48< (g-r)_0 < 0.55$.  This could give the appearance of a giant branch
at $(g-r)_0 \sim 0.5$, even if there is not one there.  However, we think most of the ``on-stream" stars
on the giant branch in Figure 8 are stream stars, because they shift slightly with $g_{\rm corr}$,
as is expected from a real giant branch, and they have a very different distribution from the
stars that were selected with all of the same criteria except the removal of the velocity cut.
Note that there are many more on-stream spectra (87 stars) than off-stream spectra (28 stars) in
the final samples, even though the original star samples were the same size.  If we exclude BHB stars
from the on-stream sample, then there are still 51 non-BHB stars in the on-stream sample compared to 
24 non-BHB stars in the off-stream sample.  This suggests that about half of the spectra of giant
stars, blue stragglers, and asymptotic giant branch stars are bona fide stream members.

In order to estimate the metallicity of Orphan Stream stars, we overlay
fiducial sequences of globular clusters M92 ([Fe/H] = -2.3), M3 ([Fe/H] = -1.57), and M13 ([Fe/H] = -1.54) 
on the color-magnitude diagram in Figure 8. 
These three clusters have blue horizontal branch stars at $g_0=15.59, 15.10$, and $14.80$,
respectively \citep{clemetal08,anetal08}.  Using the distance modulus from \citet{h96}, the inferred
absolute magnitude of the BHB stars is $M_g=0.50$ for M3, $M_g=0.52$ for M92, and
$M_g=0.38$ for M13.  These three clusters have been shifted in apparent magnitude so that
their horizontal branches align with $g_{\rm corr} = 17.75$.
The relatively metal rich cluster M71 ([Fe/H] = -0.73) has also been 
superimposed.  Since it does not have a BHB, it has been shifted so that if it did have a
BHB at the average absolute magnitude of $M_{g_0}=0.45$, at its 
cataloged $(m-M)_0 = 13.02$ \citep{h96}, then that horizontal branch would be at $g_{\rm corr}=17.75$.

Since the turnoff color and giant branch stars of our Orphan stream candidates are equally 
consistent with the M3, M92 and M13 fiducials at our level of accuracy, we
have decided to use an intermediate horizontal branch absolute magnitude of $M_{g_0}=0.45$ 
to compute distances in this paper.  This is the same as
the value used in \citet{nyw09} to measure the distance to the Cetus Polar Stream in the
Galactic halo.  Since the horizontal branch of the Orphan Stream is at $g_0=17.75$ at 
$\Lambda_{\rm Orphan}=0^\circ$, the implied distance modulus there is $\mu=17.30$; this is the
distance modulus to which all stream stars in the CMDs are shifted.  All of the distances 
in Table 2 have been computed assuming the BHBs have $M_{g_0}=0.45$.  The systematic error
in our absolute magnitude could be easily as large as 0.3 magnitudes, which is 15\% in
distance.  A systematic error in the absolute magnitudes of Orphan Stream BHB stars 
would shift all of the distances closer or farther away.  However, since the random 
errors are the important factor for fitting the Galactic potential, and not the 
systematic errors, no systematic errors have been included in the distance errors 
in Table 2.  

The turnoff, giant, and BHB stars in our spectroscopic sample could reasonably fit an
isochrone between those of M92 and M3 $(-2.3 < \rm [Fe/H] < -1.6)$, which is consistent with the
measured metallicities of the BHB stars.  However, since we do not spectroscopically sample 
very many stars that are redder than $(g-r)_0=0.55$, in either the SDSS
or in SEGUE-I, we are not sensitive to the giant branches of higher metallicity populations, even
if we include stars with higher measured metallicities in the selection.  Since we expect that
the progenitor of the Orphan Stream was a dwarf galaxy, it is reasonable that there may be
a range of metallicities in the turnoff and giant branch even if there is not a range of metallicities
in the BHB stars \citep{mswo03,chouetal07}.  

As a final test of our position, velocity, and distance selection, we show in Figure 9 the Galactic 
coordinates of spectroscopically selected BHBs.  We select BHBs from all SDSS/SEGUE
stellar spectra with the criteria $-0.4<(g-r)_0<0.3$, $0.8<(u-g)_0<1.6$, and $0.1<\log \rm g<3.5$.
We then selected all of the BHBs with apparent magnitudes consistent with BHBs at the distance of
the stream, as a function of $\Lambda_{\rm Orphan}$ ($17.4< g_0 -0.00022 \Lambda_{\rm Orphan}^2+0.034
\Lambda_{\rm Orphan} < 18.0$).  This selection really only applies within a few degrees of the
position of the Orphan Stream, but using this selection criterion over our whole dataset allows
us to visualize the density of spheroid BHBs and determine whether there is an overdensity of
BHBs with the properties of the Orphan Stream at the expected locations in the sky.  

The last two criteria for selecting BHBs in the Orphan Stream are velocity and metallicity, which
are not possible to use as photometric target selection criteria.  The BHBs with the correct
velocity to be in the Orphan Stream 
($-0.0445 \Lambda_{\rm Orphan}^2 - 0.935 \Lambda_{\rm Orphan} +95 < v_{\rm gsr} <
-0.0445 \Lambda_{\rm Orphan}^2 - 0.935 \Lambda_{\rm Orphan} + 165$)
and with low metallicity ([Fe/H]$_{\rm WBG}<-1.6$) are shown as filled circles in Figure 9.
The larger triangles show the positions of the Orphan Stream, as determined from photometry of
F turnoff stars in Figure 4 and presented in Table 2.

Our spectroscopic sample of BHBs is not uniformly sampled in magnitude, color, or position on the
sky, so there could be many selection effects in those variables.  The open circles in Figure 9
give us an impression of the density of BHB stars sampled in our color and magnitude range across
the sky.  The BHBs with the correct velocities and metallicities to be stream stars are
more abundant along the path of the Orphan Stream, which supports strongly our selection criteria, and our identification of stream members.

Note that in Figure 9 we have possible detections of the Orphan Stream at Galactic longitudes
of $l=160^\circ$ or less, suggesting that the stream does extend further than we have been
able to trace it.  Also, there is some confusion about the position of the stream at
$l=270^\circ$.  The selection of that position from photometry of F turnoff stars was very
ambiguous, and the selected point is more than two degrees from the nominal $B_{\rm Orphan}=0^\circ$
expected position.  The highest density of BHB spectra at the expected magnitude, velocity, and
metallicity for stream membership is much closer to $B_{\rm Orphan}=0^\circ$, but there are no
spectra at lower Galactic latitudes so there is an ambiguity between the spectroscopic position
at higher Galactic latitude and a broader stream that is centered at lower Galactic latitude.
We have argued that the Orphan Stream progenitor, or its remnant, is likely to be near
$l=253^\circ$.  That means there
is a reasonable chance that the stream detection at $l=270^\circ$ is part of the trailing tidal
tail, while all of the other detections are part of the leading tidal tail.  If this is the case,
then it would be reasonable that the orbit fit would not pass exactly through the detections. 

\section{Proper motions of Orphan Stream stars}

We use DR7 proper motions, which are derived from an astrometric 
match between SDSS and UCAC catalogs \citep{metal04}, as an additional check on consistency of
the proposed fit to the stream.
DR7 proper motions have typical errors of 3 $\rm mas~yr^{-1}$ per 
coordinate, restricting the usefulness to those pieces of the 
stream with $d < 25 $ kpc, in the area of Figure 1 with $230^\circ < l < 275^\circ$. 
Figure 10 shows the SDSS-UCAC proper motions of two sets of stream candidates:
1) spectroscopic stream candidates, with metallicity, 
velocity, gravity, matching the stream (all the magenta crosses from 
Figure 8);  and 2) spectroscopic BHB stream candidates 
(all the black $+$ signs from Figure 8), our most secure stream members.
Later, we will fit an orbit to the Orphan Stream, and will show that the orbit
plausibly fits this proper motion data.

\section{Metallicity of the Orphan Stream}

Throughout this paper we have been identifying Orphan Stream stars by selecting low metallicity
([Fe/H]$<-1.6$) stars.  In this section we present the statistics of stream stars selected 
using all of the position, velocity, surface gravity, magnitude, and color information 
available to us, but not using metallicity selection.

There are 36 spectroscopically selected Orphan Stream BHB stars shown in green in Figure 6
and as `+' asterisks in Figure 8.  If we use exactly the same selection criteria except omitting
the low metallicity cut, we find a sample of 39 stars.  The metallicities of the 37 of these stars 
which have measured metallicities are
histogramed in Figure 11, showing a very low metallicity peak at about [Fe/H]$_{\rm WBG}=-2.1$.
We compared these 37 stars with the 384 halo BHB stars 
in Figure 17 of \citet{ynetal09b}, which have a mean of [Fe/H]$_{\rm WBG}=-1.6$, using a
Mann-Whitney test.  The sum of the ranks of the 37 Orphan stream BHB stars is 4094, with a 
z-score of 5.25, and a p-value less than 0.0001.  The metallicities of the Orphan stream stars do not match
those of the spheroid at an extremely significant level.

In Figure 11, we also show the metallicity distribution of all of the on-stream spectroscopic 
stars in Figure 8, excluding BHB stars and bright stars with $g_{\rm corr}<16$, and all of 
the off-stream spectroscopic stars in Figure 8, also excluding BHB stars and bright stars.  The 
on-stream stars are primarily giant branch stars, but include a few
candidate blue straggler and asymptotic giant branch stars.  
There are 87 on-stream stars with spectra in Figure 8.  If we remove the low metallicity cut,
there are 136 stars, of which 39 are BHBs and 3 are bright, leaving 96 stars whose metallicity
(using the average SDSS feha metallicity measurement) is histogramed in the lower panel of
Figure 11.  In comparison, there are only 28 off-stream stars in Figure 8, and if the metallicity
cut is removed the number of off-stream stars rises to 73.  After removing bright stars with 
$g_{\rm corr}<16$, and stars that have magnitudes, surface gravities, and colors of BHB stars,
there are 59 stars left whose metallicity is histogramed in Figure 11.  Recall that the size
of the off-stream region was adjusted so that there were a similar number of SDSS/SEGUE
spectra in the on-stream region and the off-stream region.  Because coverage and target selection
criteria change as a function of sky position, it is possible that the normalization of the
on-stream and offstream datasets may not be identical or that there are slightly different
selection functions for each.  

There was no significant difference between the metallicity distribution of the on-stream
and off-stream stars in the lower panel of Figure 11, using either a Mann-Whitney test
or a KS test.  Both samples have a mean metallicity of [Fe/H] $\sim -1.6$.  However, it appears 
to the eye that there are a few more very low metallicity
stars in the off-stream sample and a few extra stars at about [Fe/H]$=-1.9$ in the on-stream
sample.  The difficulty here is that we are using very small samples, and 50\% of the on-stream
stars are likely contamination from background stars, so better and cleaner samples
are required to study the metallicity distribution of the population as a whole.
It is not unexpected for BHB stars, which only form from low metallicity populations,
to be more metal-poor than giant branch stars.   

\section{Orbit Fitting}

\subsection{Methodology and Model Choices}

We now use the combined positions, distance estimates, and line-of-sight velocities to fit an 
orbit to the Orphan Stream.
We use the same procedure described by \cite{wetal09} with two modifications. First, instead 
of fitting the distance to the stream, we fit the $g_0$ magnitude. We do this because of the 
symmetry of the errors in $g_0$. Specifically we adopt a reduced chi-squared of

\begin{equation}
\chi^2_{b} = \sum_{i} \left(\frac{b_{model,i} - b_{data,i}}{\sigma_b}\right)^2,
\label{bchisq}
\end{equation}

\begin{equation}
\chi^2_{rv} = \sum_{i} \left(\frac{rv_{model,i} - rv_{data,i}}{\sigma_{rv,i}}\right)^2,
\label{vchisq}
\end{equation}

\begin{equation}
\chi^2_{g_0} = \sum_{i} \left(\frac{g_{0model,i} - g_{0data,i}}{\sigma_{g_0,i}}\right)^2, {\rm ~and}
\label{gchisq}
\end{equation}

\begin{equation}
\chi^2 = \frac{1}{\eta} \left(\chi^2_{b} + \chi^2_{v} + \chi^2_{g_0}\right),
\label{chisq}
\end{equation}
where $\eta = N - n -1$, $N$ is the number of data points,
and $n$ is the number of parameters.
Distances for the fiducial stream points are calculated from the $g_0$ magnitudes for BHBs in 
Figure 7 under the assumption that the absolute magnitude for a stream BHB is 
$M_{g0} = 0.45$ at about $(g-r)_0 = -0.15$, similar to
the globular cluster fiducials overlaid in Figure 8.  The distance parameter is fitted
internally in the gradient descent search, but  
because the error bar as 
represented in distance space is not symmetric above and below the fitted value,
the chi-squared computation is done in magnitude space.

Second, in addition to the logarithmic halo potential model, we also consider an NFW \citep{nfw96}
model.  Although our data does not span a large enough range of Galactocentric distances
to determine which model is preferred, it is useful to optimize both models for easy
comparison with the literature.

Likewise, we will use two published disk potentials: a lower mass exponential disk from 
\citet{xetal08} and a higher mass Miyamoto-Nagai disk with parameters from \citet{ljm05}.
While we do not have sufficient data to constrain the disk model, it is instructive to see
the effect of using different, commonly used disk models on the inferred halo mass.

To obtain our initial guess for the kinematic parameters, we draw from Tables 1 and 2 a test particle at 
Point 1: $(l,b,R) = (218^\circ, 53.5^\circ, 30 \rm ~kpc)$ and construct a vector to 
Point 2: $(l,b,R) = (215^\circ, 54.0^\circ, 31 \rm ~kpc)$. By matching the radial velocity at Point 1, we obtain an initial parameter starting point of $(R,v_x,v_y,v_z) = (30 \rm ~kpc, -125 \rm ~km \rm ~s^{-1}, 75 \rm ~km \rm ~s^{-1}, 95 \rm ~km \rm ~s^{-1})$.

With this initial kinematic guess we will consider seven specific cases, which include three potential models
published by previous authors, and four models in which the halo mass is varied to best match our data
(logarithmic and NFW halos are compared, as well as a low mass exponential disk vs. a high mass
M-N disk): 

\begin{enumerate}
\item We fit the exact model of \citep{xetal08}, using an exponential disk and
NFW halo.  The parameters for bulge, disk, and halo were taken from their paper.
the mass of the bulge, disk and halo, integrated out to 60 kpc, are tabulated in
Table 3.  The mass of the Milky Way using this model is $M(R<60\rm ~kpc) = 4.0\times 10^{11} M_\odot$, of
which $3.3\times 10^{11} M_\odot$ is in the halo.
The value for the scale radius, $r_s = 22.25 \rm ~kpc$, used for all the NFW potentials in this work 
comes from the top panel of Figure 16 of \citet{xetal08}.  For a Milky Way virial radius of
$r_{\rm vir} = 240 \rm ~kpc$ \citep{2007MNRAS.379.1475S} this corresponds to 
a concentration index of $c\sim 11$.

\item We fit the same model as the previous case, but allow the $v_{\rm c,max}$ 
normalization of the NFW model (the halo mass) to vary in amplitude.

\item We fit the same bulge and exponential disk model as the previous two cases, but use a 
logarithmic halo model.  We allow the normalization (mass) of the halo to vary as a free 
parameter to be fit.

\item We fit stream kinematics only, assuming a spherical logarithmic halo as given 
by \citet{ljm05}.  Disk and bulge parameters for this case are identical to the \citet{ljm05}
paper.  Their model, with a disk model from \citet{1975PASJ...27..533M},
yields $M(R <60\rm ~kpc) = 4.7\times 10^{11} M_\odot$.

\item 
We fit the same model as case 4, but allow the halo speed above to vary as a 
free parameter within the spherical logarithmic halo potential model, while 
keeping $d = 12 \rm ~kpc$, where $d$ is the halo core softening radius.

\item We fit the kinematics within the spherical NFW halo given in \cite{nfw96} and further 
described by \cite{kgkk99}, normalized to $M(R<60\rm ~kpc) = 3.3\times 10^{11} M_\odot$. 
We fit the $v_{c,max}$ of the NFW model to give a similar total potential
to that for the logarithmic halo with $v_{\rm halo} = 114 ~\rm km~s^{-1}$ above.
These give rise to $v_{c,max} = 155 \rm ~km \rm ~s^{-1}$ and $r_s = 22.25 \rm ~kpc$.
\item We fit the same model as case 6, but allow the NFW maximum circular speed $v_{c,max}$ to vary while keeping $r_s = 22.25 \rm ~kpc$.

\end{enumerate}

We summarize these potential models in Table 3.

The models were fit to all of the available data in Tables 1 \& 2, with the exception of
the last point in Table 2.
Attempting to fit the data set with the $l = 271^\circ$ point included resulted in 
substantially worse $\chi^2$ values were obtained ($\chi^2 \sim 4$).  Additionally, fitting this point introduced deviations of the orbit from the observed stream 
distances, with the high $l$ distances being underestimated, and the low $l$ distances 
being overestimated. This systematic differences between the model and data led us to 
perform all of our best fit models without the $l = 271^\circ$ point, for all six of the 
orbit fits listed above.

The datapoint at $l=271^\circ$ is less certain because the measurements were
problematic, and because the continuity of the 
Orphan orbit is broken by the missing SDSS
data at $255^\circ < l < 268^\circ$, and the location of a possible Orphan progenitor
dwarf in this region, while likely, is uncertain. If the progenitor is in this
region, then the points with $l < 255^\circ$ are in a tidal tail with the opposite
sense (i.e. leading) to those with $l > 268^\circ$ (i.e. trailing). If this
is the case, than a simple orbit fit, (rather than a full N-body simulation), will
not accurately account for the deviations in distance and velocity that
occur on opposite sites of a progenitor when that progenitor has an internal
velocity dispersion that is a significant fraction (more than a few percent) of the
total potential in question.

In addition to these cases, we also attempted several more which did not produce very interesting
results, but which we will summarize here.  We ran several orbits fits in an attempt to
measure the halo flattening parameter, $q$.  The
fits were very insensitive to $q$; fitting this parameter 
in the logarithmic halo yielded results with a very large error in $q$.  Therefore,
we fixed the flattening parameter at $q=1$.

We also attempted to simultaneously fit $v_{\rm halo}$ and $d$, and $v_{c,max}$ and $r_s$,
for the logarithmic and NFW halos, respectively. In general, fitting the 
scale lengths did not substantially change the halo speed results, and the scale 
lengths were fit only with very large errors, on the order of tens of 
kiloparsecs.   Therefore, we did not attempt to fit the scale lengths in either model.

\subsection{Results}

For each halo model, we ran five random starts of the fitting algorithm in the same manner 
described by \cite{wetal09}. Once the best fit orbit was found for each model, we computed
the Hessian errors by varying the parameter step sizes until the errors converged for different 
step size choices. Using a starting point 
of $(l,b) = (218^\circ, 53.5^\circ)$, the best fit parameters and their errors are 
enumerated in Table 3.

Figure 12 shows the three orbit fits for the models with the exponential disk.
The position and velocities of the datapoints are well fit by all three exponential
disk models.
The lower panel of Figure 12 shows that for this stream, the distance fit is most sensitive to the
potential.  The weaker potential of the lower mass halo model 
gives the best fit to observed stream distances from Tables 
1 and 2 (black points with error bars), as it allows the stream to 
escape to larger Galactocentric radii
without significantly changing the observed radial velocity gradient.
There is little difference between the best fit NFW halo and the best fit logarithmic
halo.

Figure 13 shows the circular velocity curve for models N=1-3 from Table 3, which use
the same exponential disk parameters as \citep{xetal08} in all cases.
The figure shows that the orbits that are well fit to the Orphan Stream don't fit 
the \citep{xetal08} nor the \citep{krh09} result; the rotation speed of the model is
too low at all Galactic radii.

We therefore try using a M-N disk with parameters as adopted by \citep{ljm05}.  This
disk is about twice massive than the exponential disk model.  We note here that we have not
fully explored all possible disk shapes and masses, and a more massive exponential
disk may give similar results to the more massive M-N disk considered here.
Figure 14 shows the four fits for the models with the M-N disk.
As before, lower halo speeds are a better fit to the Orphan Stream than those used by 
previous authors, and there is little difference between the fits for the best fit NFW and
logarithmic halos.
Evidently, the Orphan Stream is telling us that the mass of the Milky Way is smaller
than previously thought, so that distant portions of the stream, which are therefore experiencing lower
gravitational attraction to the Galaxy, are pushed further away from
the Galaxy center at a given energy.

Figure 15 shows the same circular velocity as Figure 13 for models 4-7 with an M-N disk.
This time we find that the low mass halo orbit fits {\it are} a good fit to the \citet{xetal08}
and \citep{krh09} result.  They actually fit the \citet{xetal08} model better than the
model fit in that paper.
Using the higher velocity leads to a systematic deviation in the distances, so
that the model is further away than all of the datapoints at high Galactic latitude, and
closer for low Galactic latitudes.  Also, as we will see later, adding velocity segregation
from an N-body simulation only makes the systematic errors in the distance larger;
for example stars in the leading tidal tail have lower total energy than the progenitor and have smaller
Galactocentric distances.  The energy difference increases as a function of distance
from the progenitor, along the stream.  However, we note that our formal error bar on 
$v_{\rm halo} = 73 \pm 24 \rm ~km~s^{-1}$ is quite large and marginally consistent 
(within 2 sigma) with the 
higher $v_{\rm halo} = 114 \rm ~km~s^{-1}$ value of \cite{ljm05} and others.
Although the formal error bars are large, the orbit fits suggest that a lower
halo speed is preferred.

Of the models that we tried, the three low-mass exponential disk models (1-3) fit either
the Orphan Stream data or the \citet{xetal08} rotation curve, but not both.  The two
low halo mass models (models 5,7) with a higher mass M-N disk fit the best, with a logarithmic
halo fitting about the same as an NFW profile.  The two higher halo mass models (models 4,6)
are poorer fits to both the Orphan Stream and the \citet{xetal08} rotation curves.

Although we cannot formally rule out the halo models fit by \citet{xetal08}
and \citet{ljm05}, in order to simultaneously fit the \citet{xetal08} circular
velocity data, we prefer a lower total mass of the Milky Way than measured in
either of those two papers.  Our total Milky Way mass is 60\% of that found
by \citet{xetal08} and \citet{ljm05}.  As mentioned in the introduction, our
masses are at the low end, but not out of range of recently published masses.
To estimate the virial mass of the Milky Way given our best fit model 5, we
multiply the halo mass within 60 kpc by 4 (for a virial radius of 240 kpc), and
add the disk plus bulge mass.  The result is $M(R<240{\rm ~kpc})=6.9 \times 10^{11} M_\odot$.
Of interest is the recent result of
\citet{oetal09}, who suggested that a lower halo mass might make it easier to fit
the kinematics of the Pal 5 globular cluster stream.
The findings on Milky Way circular velocity with radius to 
distance of $R \sim 60$ kpc by \citet{xetal08} also are consistent 
with our low value for the halo speed.  When the M-N disk model
is used, they are also in rough agreement with \citet{krh09} who measure
$v_c(R=8\rm kpc) = 224 \pm 13 ~\rm km~s^{-1}$. 

We note that we didn't explore scaling the mass of the disk (or bulge) beyond 
their default values of $M_{\rm disk} = 10^{11} M_\odot$ for the M-N disk or
$M_{\rm disk} = 5\times 10^{10}$ for the exponential disk.  Thus, while
we find that the heavier M-N disk (combined with a lower amplitude halo) 
fits the joint data sets of the Orphan Stream, and the \citet{xetal08} BHBs 
circular velocity points, better than models with the lighter exponential disk, it 
is likely that similar fits could be obtained with a heavier exponential disk (though 
still with a lighter halo).  If more data were available, the disk mass could be
varied to find the best combination of disk and halo mass.

It is important to note from Figures 12 and 14 that if we could follow the Orphan
Stream just a little farther out into the halo, we would have a much better power
to determine the halo mass, since the distances to the stream for each case diverge.

Figure 16 shows the best fit $v_{\rm halo} = 73 \rm ~km \rm ~s^{-1}$ logarithmic halo model in 
(X,Y,Z) Galactic coordinates with the direction of motion indicated by the arrows.  
From the apo- and peri-galactic distances of $\sim 90 \rm ~ kpc$ and $\sim 16.4 \rm ~kpc$, 
respectively, we calculate an orbital eccentricity of $e = 0.7$. The inclination of the 
Orphan Stream is $i \sim 34^\circ$ with respect to the Galactic plane as seen from the 
Galactic center.  It will be important to obtain data for the Orphan stream at apogalacticon
in order to distinguish between different models for the Galactic potential.

The proper motions along the model orbits of Figure 14 are calculated 
for the three potentials.  The implied proper motions are essentially identical in
the region of the stream where it is observed $170^\circ < l < 270^\circ$, and thus only the
best fit model (model 5) is overlaid as a solid curve on Figure 10.  
The fact that the stream is moving in prograde orbit (see below) makes the proper motions 
significantly smaller in magnitude than if the stream were retrograde \citep{wetal09},
and thus, while the errors on the proper motions and stream membership 
uncertainties make it difficult to constrain the Galactic potential 
with proper motions of this accuracy, it is reassuring that the data and
model are broadly consistent.

\subsection{The Orphan Stream is moving in a prograde direction}

The velocities of the fit orbit definitively show that the stream stars 
are moving in a direction from higher $l$ to lower $l$.  As a simple check, if 
we reverse the sign of all the velocities ($v_x, v_y, v_z$) in the model orbit, 
the path on the sky that the stream traces out is the same, but the radial velocities 
do not match the observations by a wide amount ($|\delta(v)|> 100~ \rm km~s^{-1})$.  
Thus the Orphan Stream is on a prograde orbit around the Milky Way.
One may now ask if the visible piece of the Orphan tidal stream is a leading or
trailing tidal tail.  If we assume that the `progenitor' of the Orphan Stream lies
in the range $248^\circ < l < 268^\circ$ as the density plot of Figure 5 and the visual
impression of Figure 4 suggests, (with the density enhancement visible
 at $(l,b) = (253^\circ,49^\circ)$), then the portion of the tidal stream 
stretching from $l=250^\circ$ to $l=170^\circ$, combined with the space velocity 
of stars moving in that same direction, implies that we are seeing a 
leading tidal arm, rather than a trailing tidal arm.   The piece of stream
intercepted at $l=270^\circ$ is then a trailing tidal arm.

\subsection{N-Body Realization}

We now comment on the fact that our `simple orbit integration and fitting' technique 
is not an N-body simulation, and therefore effects of stars leaving the 
progenitor and drifting ahead or behind to greater 
and lower energies, which changes their distance from the central body, 
are not captured as they would be in a full N-body.

To conduct the N-body simulation, we first integrate the log halo (best fit model 5) orbit 
back $4 \rm ~Gyr$ and place a 10,000 particle Plummer sphere at the predicted location. We 
use a Plummer sphere scale radius of $r_s = 0.2 \rm ~kpc$ and total mass of 
$M_{\rm total} \sim 2.5 \times 10^6 M_\odot$. Using the orbit kinematics from $4 \rm ~Gyr$ in the past, we evolve it forward for $3.945 \rm ~Gyr$ so that the dwarf progenitor ends up at $l \sim 250^\circ$. We conducted the N-body simulation using the gyrfalcON tool of the NEMO Stellar Dynamics Toolbox \citep{teuben}. 

The result of this simulation is given in Figure 17. The top three panels show the logarithmic halo orbit with $v_{\rm halo} = 73 \rm ~km \rm ~s^{-1}$ and N-body results for $(l,b), ~(l,v_{\rm gsr}), ~\rm and ~ (l,d_\odot)$.   

We see that the N-body stream is a good fit to the sky locations and
velocities of the simple orbit (black curve in Figure 17).  In the 
distance comparison (third panel), the N-body is significantly below the orbit
at low $l$.  This is in agreement with our intuition for a stream that is moving
from high $l$ to lower $l$, since energy differentials due to the internal 
dispersion of stars in the progenitor draw stars with lower energy (and lower
Galactocentric distance) further ahead closer to the Galactic center than those with 
higher total energy.  Note that this is independent of the question of whether
the tail is leading or trailing.  We know the direction that the stream is going
from the observed radial velocities and distances.  We expect that we are seeing
a leading tidal tail because we think the dwarf remnant is located at the high Galactic
longitude end of the observed portion of the stream.  But even if we are incorrect
about the position of the dwarf, the observed slopes in the lower panels of figures 
12 and 14 indicate a lower mass Milky Way, and a full N-body simulation will only
make the slope for higher mass estimates fit worse.

The lowest panel in Figure 17 shows the density of N-body simulated particles remaining 
at the current epoch along the stream vs. $l$.
This density distribution may be compared with that of the data in Figure 5.
The N-body simulation has a lower density of particles 
near $l = 240^\circ, ~(\Lambda_{\rm Orphan} = 15^\circ)$.  These two effects are apparent 
in the Orphan imaging data in Figures 5 and 4, respectively. 
This suggests that our method of fitting a simple orbit to the tidal debris is
a reasonable first approximation, and fitting the density is a useful diagnostic tool.

\section{Connection of the Orphan Stream to other Galactic halo structures}

Given the Orphan Stream orbit of \S 10, we can revist the question the Orphan Stream's connection to other halo structures, such as Segue-1, Complex A and Ursa Major II.
We add position, velocity, and distance using the data 
from the literature for these three objects as points onto 
Figures 12 and 14 and 16 for reference.

\subsection{Segue-1}

The distance to Segue-1 is $d=23$ kpc from
the Sun \citep{2007ApJ...654..897B}, consistent with both Sagittarius and Orphan Streams in this part of the sky.
A case has recently been made that Segue-1 is associated with the 
Sagittarius stream \citep{noetal09}. It is, however, in the same general 
area of the sky as the Orphan Stream, and it has a 
positive velocity, $v_{\rm gsr} = +114 \rm~km~s^{-1}$, as measured 
by \citet{getal09}, very close to that of Orphan Stream stars in Table 2 
here, and in contrast to the Sgr stars in this area, 
which have $v_{\rm gsr} \sim -120 \rm~km~s^{-1}$, as shown in \citet{ynetal09b}.  
Thus, on the basis of kinematics, we believe that it is 
more likely that Segue-1 is a star cluster associated with the Orphan Stream than
that it is associated with Sgr.  More extensive proper motions of 
both objects will help resolve this issue definitively \citep{getal09}.

\subsection{Ursa Major II}

In Figure 1, Ursa Major II \citep{zetal06} is visible at the right hand side 
of the diagram near $(l,b) = (152^\circ,37^\circ)$, not far from the extension of the
projected Sgr orbit.
Kinematics of UMa II \citep{metal07}, show it 
to have a $v_{\rm radial} = -115 \rm~km~s^{-1}$ 
(following the body and figures of that paper rather than the abstract) which corresponds 
to $v_{\rm gsr} \sim -35\rm ~km~s^{-1}$.  This is in strong disagreement with the Orphan 
model predictions of $v_{\rm gsr} \sim +60$ km s$^{-1}$ at the approximate location of
UMa II.  The distance $d_\odot$ to the object is also in disagreement
with the models in Figures 12 and 13.  Additionally, the density of Orphan Stream stars, were
UMa II the Orphan progenitor, should increase as one gets closer to UMa II, not 
decrease.  For these reasons, given the improved SEGUE data points 
and implied Orphan Stream kinematics, the current data do not support 
the hypothesis that UMa II is the progenitor of the Orphan Stream.

\subsection{Complex A}
As discussed above, and in \citet{setal09}, Complex A is much too close to 
be associated with the Orphan Stream without extensive gas dissipation.
It would take a strongly dissipative interaction
and a significant amount of time for the HI gas to become stripped from the Orphan Stream and migrate to lower Galactocentric distances.

\subsection{Other possible associations}

We check our best fit orbits against known globular clusters and dwarf galaxies to see if there are any possible associations. Of the globular clusters given in \cite{h96}, we find six that lie within 5 degrees of the orbits: NGC 6752, NGC 6362, Rup 106, Pal 1, NGC 7078, and NGC 7089. Of these, only three have radial velocities consistent with the orbit: NGC 6362, NGC 7078, and NGC 7089. However, these are approximately 10 kiloparsecs from the Sun, while the orbit is consistently above 20 kiloparsecs. Therefore we rule out these as being associated with this stream.

In addition to Ursa Major II, we find three additional dwarf galaxies that lie within 5 degrees of the projected orbits: Leo II, Leo A, and Ursa Major Y. Of these, only Leo A has a radial velocity that could associate it with the orbit, but its distance of $800 \rm ~kpc$ from the Sun precludes this possibility.

\section{Conclusions}

We present substantially better spatial and radial velocity measurements of
Orphan Stream stars than previous papers, and are therefore able to fit a
better orbit solution, which is substantially different from previous papers.

While the Orphan Stream was so named because of the absence of a progenitor, 
there is a suggestion from the data in this paper that the progenitor is 
located near $(l,b) = (253^\circ,49^\circ)$, right at the edge of 
the SDSS survey footprint.  
Like previous papers, our data suggest a small dwarf galaxy to be the likely 
progenitor, since the velocity dispersion of the stream stars is $\sim 10 \rm ~km~s^{-1}$.

We summarize our findings as follows:

1) The Orphan Stream is visible on the sky in SDSS imaging and traces an arc
from $(l,b) = (271^\circ, 38^\circ)$ to $(l,b) = (171^\circ,46^\circ)$ with a $v_{\rm gsr} \sim +110\pm 20 ~\rm km~s^{-1}$ for most of this length at distances increasing from
$d_\odot = 18\rm ~kpc$ to over $d_\odot = 46 ~\rm kpc$, where it is no longer detectable
in our data.
The stream is on a prograde orbit with perigalacticon, apogalacticon, eccentricity and
inclination of $16.4 \rm ~kpc, \sim 90 ~kpc, 0.7, \rm ~and~ 34^\circ$, respectively.

2) We identify F turnoff stars and BHBs in the Orphan Stream, along with candidate 
blue stragglers.
All of these populations except F turnoff stars also have representatives with SEGUE spectra;
there may also be F turnoff stars with spectra, but they are so faint that surface
gravities could not be measured.  The turnoff color is $(g-r)_0=0.22$.

3) The metallicity of the Orphan Stream BHBs is
very low, [Fe/H] $= -2.1\pm 0.1$.

4) The progenitor for 
the Orphan Stream may be located between $(l,b) = (250^\circ,50^\circ)$ 
and $(270^\circ,40^\circ)$. This assumption, combined with the 
observation that the Orphan Stream is on a prograde orbit 
around the Galaxy implies that the tidal tail at $l < 250^\circ$ is a leading tidal tail.

5) The Orphan Stream data is best fit to a Milky Way potential with a halo plus disk plus
bulge mass of about $2.6 \times 10^{11} M_\odot$, integrated to 60 kpc from the Galactic center.  
Our best fit is found with a logarithmic 
halo speed of $v_{\rm halo}=73\pm 24~ \rm km~s^{-1}$, a disk+bulge mass of
$M(R< 60 {\rm ~kpc}) = 1.3 \times 10^{11} M_\odot$, and a halo mass of
$M(R< 60 {\rm ~kpc}) = 1.4 \times 10^{11} M_\odot$.  
However, we can find similar fits to the data that use an NFW halo profile.  Our halo speed of 
$v_{\rm halo}=73\pm 24~ \rm km~s^{-1}$ is smaller than that of previous literature.
Although our fits with smaller disk masses and
correspondingly larger halo masses are not good fits the the \citet{xetal08} rotation
curves, we have not tried enough models to rule out this possibility.
Distinguishing between different classes of models
requires data over a larger range of distances.  

6)The Orphan Stream is projected to extend 
to 90 kpc from the Galactic center, and measurements of these distant parts of the stream 
would be a powerful probe of the mass of the Milky Way.

7) An N-body simulation yields an excellent fit to the observed data, including matching
the approximate shape of the stellar stream star density along the stream.  The N-body  
mass used was about $M_{\rm total,Orphan} = 10^6 M_\odot$, about $10^{-3}$ the 
total mass of the Sgr stream and dSph system.  The total Orphan system mass 
is not highly constrained.

8) The list of possible halo objects associated with the Orphan Stream is reduced to one:
The `dissolved star cluster' Segue-1.  Other possible associated objects, namely
UMa II, the Complex A HI cloud, and the halo's globular clusters are not close 
to the Orphan Stream orbit presented here in distance or velocity (or both).  

Because the orbit fit is 
not able to significantly constrain the flattening $q$ of the halo potential,
we assume a $q=1.0$ throughout.
Nevertheless, the analysis of the Orphan Stream shows us that we are 
able to constrain another important halo potential parameter, namely the amplitude 
of the halo potential.  
A spherical logarithmic potential with 60\% of the mass of the fit of 
\citet{xetal08} provides a reasonable fit to 
the Orphan Stream data.  

\acknowledgments 

We acknowledge useful discussions with Steve Kent.  We acknowledge helpful
comments from the referee which significantly improved the
presentation of the results of this paper.
This work was supported by the National Science Foundation, grant AST 06-06618.  
Funding for the SDSS and SDSS-II has been provided by the Alfred
P. Sloan Foundation, the Participating Institutions, the National
Science Foundation, the U.S. Department of Energy, the National
Aeronautics and Space Administration, the Japanese Monbukagakusho, the
Max Planck Society, and the Higher Education Funding Council for
England. The SDSS Web Site is http://www.sdss.org/.

\begin{deluxetable}{rrrrrrr}
\tabletypesize{\scriptsize} \tablecolumns{7} \footnotesize
\tablecaption{Orphan Stream Detections} \tablewidth{0pt}
\tablehead{
\colhead{$l$} & \colhead{$b$} & \colhead{$\delta b$} & \colhead{$\Lambda_{Orphan}$} & \colhead{$B_{Orphan}$} & \colhead{BHB $g_0$} & \colhead{$\delta g_0$} \\
\colhead{$^\circ$} & \colhead{$^\circ$} & \colhead{$^\circ$} & \colhead{$^\circ$} & \colhead{$^\circ$} & \colhead {mag} & \colhead{mag}
}
\startdata
255 & 48.5 & 0.7 & 22.34 & 0.08 & 17.1 & 0.1 \\
245 & 52.0 & 0.7 & 15.08 & 0.56 &      &     \\
235 & 53.5 & 0.7 & 8.86 & 0.21 &       &     \\
225 & 54.0 & 0.7 & 2.95 & -0.23 & 17.6 & 0.2 \\
215 & 54.0 & 0.7 & -2.93 & -0.33 & 17.9 & 0.1 \\
205 & 53.5 & 0.7 & -8.85 & -0.09 & 18.0 & 0.1 \\

195 & 52.0 & 0.7 & -15.08 & 0.05 & & \\
185 & 50.5 & 0.7 & -21.42 & 1.12 & 18.6 & 0.1 \\
175 & 47.5 & 0.7 & -28.59 & 1.88 &      &     \\
171 & 45.8 & 1.0 & -31.81 & 2.10 & & \\
\enddata
\end{deluxetable}

\begin{deluxetable}{rrrrrrrrrrrr}
\tabletypesize{\scriptsize} \tablecolumns{12} \footnotesize
\tablecaption{Orphan Stream Detections with Velocities} \tablewidth{0pt}
\tablehead{
\colhead{$\Lambda_{Orphan}$} & \colhead{$l$} & \colhead{$b$} & \colhead{$\delta b$} & \colhead{BHB $g_0$} & \colhead{$\delta g_0$} & \colhead{$v_{\rm gsr}$} & \colhead{$\delta v_{\rm gsr}$} & \colhead{$\sigma_v$} & \colhead{N} & \colhead{$d$} & \colhead{$\delta d$}  \\
\colhead{$^\circ$} & \colhead{$^\circ$} & \colhead{$^\circ$} & \colhead{$^\circ$} & \colhead {mag} & \colhead{mag} & \colhead{km s$^{-1}$} & \colhead{km s$^{-1}$} & \colhead{km s$^{-1}$} & \colhead{} & \colhead{kpc}& \colhead{kpc} 
}
\startdata
-30 & 173 & 46.5 & 0.7 & 18.8 & 0.2 & 115.5 & 6.7 & 11.5 & 4 & 46.8 & 4.5 \\
-20 & 187 & 50.0 & 1.0 & 18.5 & 0.1 & 119.7 & 6.9 & 11.9 & 4 & 40.7 & 1.9 \\
 -9 & 205 & 52.5 & 0.7 & 18.0 & 0.1 & 139.8 & 4.6 & 12.9 & 9 & 32.4 & 1.5 \\
 -1 & 218 & 53.5 & 1.0 & 17.8 & 0.1 & 131.5 & 3.1 & 8.2 & 8 & 29.5 & 1.4 \\
 8  & 234 & 53.5 & 0.7 & 17.4 & 0.1 & 111.3 & 11.1 & 11.1 & 2 & 24.5 & 1.2 \\
 18.4& 249.5 & 50.0 & 0.7 & 17.1 & 0.1 & 101.4 & 2.9 & 9.8 & 12 & 21.4 & 1.0 \\
 36 & 271 & 38.0 & 3.5 & 16.8 & 0.1 & 38.4  & 1.7 & 2.5  & 3 & 18.6 & 0.9 \\
\enddata
\end{deluxetable}

\begin{deluxetable}{rrrrrrrrrrrrrrr}
\tabletypesize{\scriptsize} \tablecolumns{15} \footnotesize
\tablecaption{Orphan Stream Models} \tablewidth{0pt}
\tablehead{
\colhead{$N$} & \colhead{$M_{\rm Bulge}$} & \colhead{Disk} & \colhead{$M_{\rm disk}$} & \colhead{Halo} & \colhead{$M_{\rm halo}$} & \colhead{$d/r_s$}  & \colhead{$R_{\rm fit}$} & \colhead{$v_x$} & \colhead{$v_y$} & \colhead{$v_z$} & \colhead{$v_{\rm halo}$} & \colhead{$\chi^2$} & \colhead{$M_{60}^{a}$}  \\
\colhead{} & \colhead{$10^{10} M_\odot$} & \colhead{type} & \colhead{$10^{10} M_\odot$} & \colhead{type} & \colhead {$10^{10} M_\odot$} & \colhead{kpc} & \colhead{kpc} & \colhead{$\rm km~s^{-1}$} &  \colhead{$\rm km~s^{-1}$} &  \colhead{$\rm km~s^{-1}$} &  \colhead{$\rm km~s^{-1}$} & \colhead{} & \colhead{$10^{10} M_\odot$} 
}
\startdata
1 &1.5&Exp&5&NFW & 33 & 22.25 & $28.8$ & $-170\pm 11$ & $94\pm 2$ & $108\pm 10$ & 155 & 1.55 & 40 \\
2 &1.5&Exp&5&NFW & 20 & 22.25 & $28.5$ & $-157\pm 10$ & $78\pm 12$ & $107\pm 9$& $120\pm 7$ & 1.37 & 24 \\
3 &1.5&Exp&5&Log & 17.6 & 12  & $28.4$ & $-152\pm 12$ & $72\pm 12$ & $106\pm 9$ & $81\pm 12$ & 1.35 & 26.5 \\
4 &3.4&M-N&10&Log & 35 & 12  & $28.9$ & $-179\pm 11$ & $106\pm 2$ & $109\pm 10$ & 114 & 1.98 & 47 \\
5 &3.4&M-N&10&Log & 14 &12  & $28.6$ & $-156\pm 10$ & $79\pm 1$ & $107\pm 9$ & $73\pm 24$ & 1.70 & 26.4 \\
6 &3.4&M-N&10&NFW & 33  & 22.25 & $28.9$ & $-178\pm 5$ & $106\pm 3$ & $108\pm 10$ & 155 & 1.96 & 43.5 \\
7 &3.4&M-N&10&NFW & 16  & 22.25 & $28.7$ & $-161\pm 11$ & $85\pm 1$ & $107\pm 9$ & $109\pm 31$ & 1.73 & 28.4 \\
\enddata
\tablenotetext{a} {Mass due to sum of bulge, disk, and halo components out to R = 60 kpc}
\end{deluxetable}

\clearpage
\setcounter{page}{1}

\begin{figure} 
\includegraphics[scale=0.75,viewport=0in 4in 6in 14in]{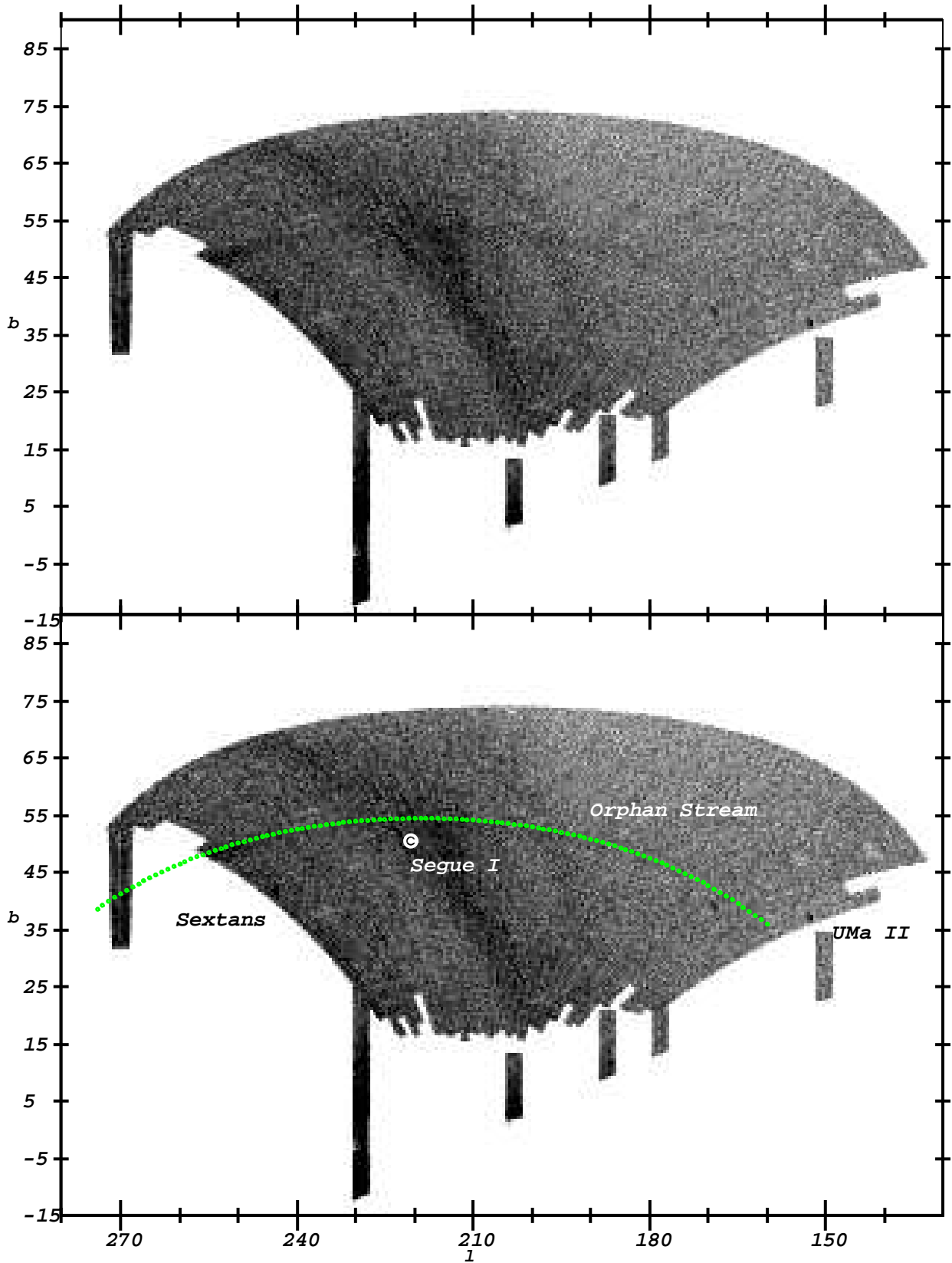}
\caption[Sky position of Orphan Stream]{
\footnotesize
We show the sky position of the Orphan Stream as traced by F turnoff stars with 
$0.12<(g-r)_0<0.26, (u-g)_0>0.4$, and $20.0<g_0<22.5$.  Darker areas of the greyscale 
plot have higher stellar density.  The most prominent overdensity, going from 
$(l,b)=(240^\circ,65^\circ)$ to $(l,b)=(200^\circ,15^\circ)$, is from the 
Sagittarius dwarf tidal stream.  The density falls off at higher latitudes because its 
stars are too faint for our selection.  Twenty degrees to the right, and running parallel to Sagittarius, is another stream that is generally referred to as a ``bifurcation" of
the Sagittarius dwarf tidal stream.  The Orphan Stream is a nearly horizontal arc in this
projection, at $b\sim50^\circ$.  There are very many stream stars at $l=255^\circ$.
The positions of the Sextans and Ursa Major II dwarf galaxies are labeled in the
lower panel.  The lower panel shows where $B_{\rm Orphan}=0^\circ$, which is our
initial fit to the sky position of the Orphan Stream.
}
\end{figure}

\begin{figure} 
\includegraphics[scale=0.65,angle=-90]{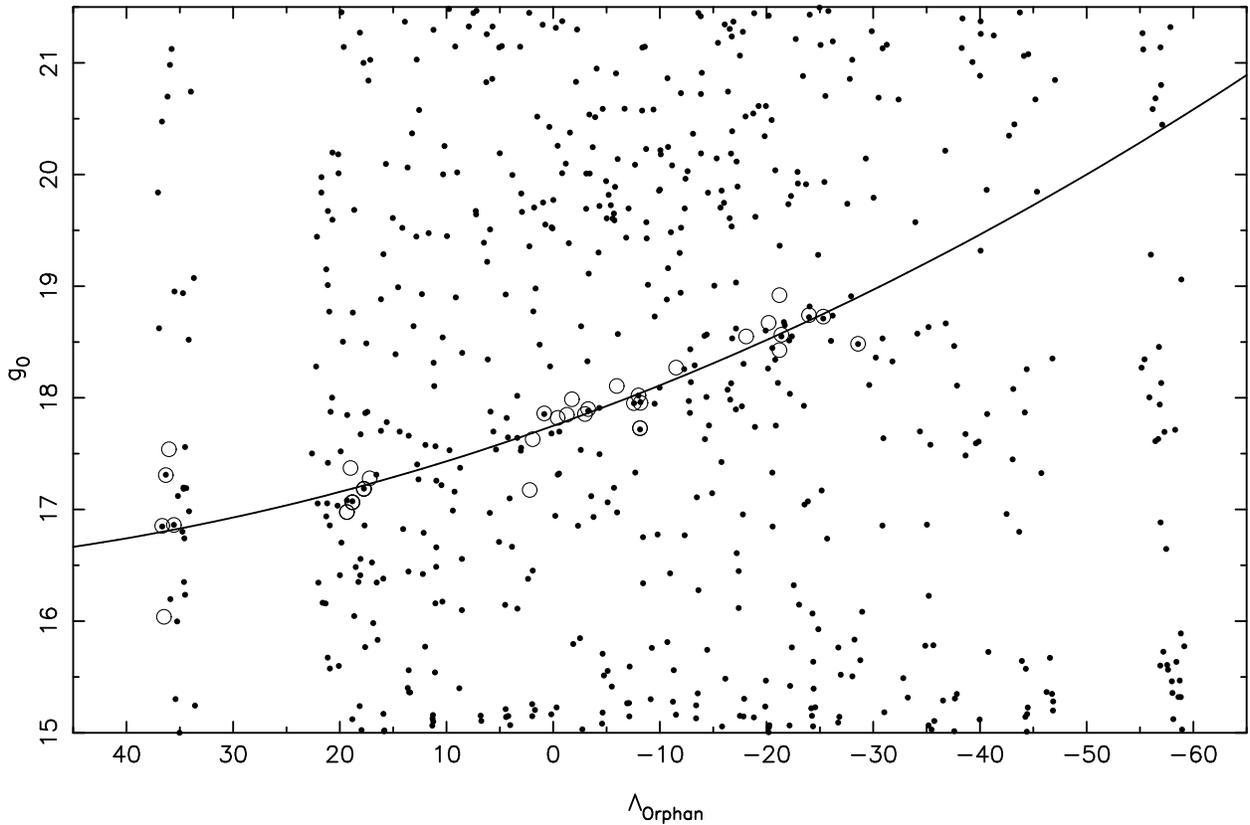}
\caption[$g_0$ vs. Lambda]{
\footnotesize
We determine the distance to the Orphan Stream from a plot of $\Lambda_{\rm Orphan}$ vs. $g_0$ for 
photometrically selected BHB stars that are within one degree of $B_{\rm Orphan}=0^\circ$.
We also show the positions of spectroscopically selected BHB stars that are within one degree
of $B_{\rm Orphan}=0^\circ$, have low metallicity ([Fe/H]$_{\rm WBG}<-1.6$), and are 
clustered in distance and velocity consistent with membership in a tidal debris stream.
The black line shows our fit to the distance of the stream as a function of $\Lambda_{\rm Orphan}$.
}
\end{figure}

\begin{figure}
\includegraphics[scale=0.75,viewport=0in 0in 6in 10in]{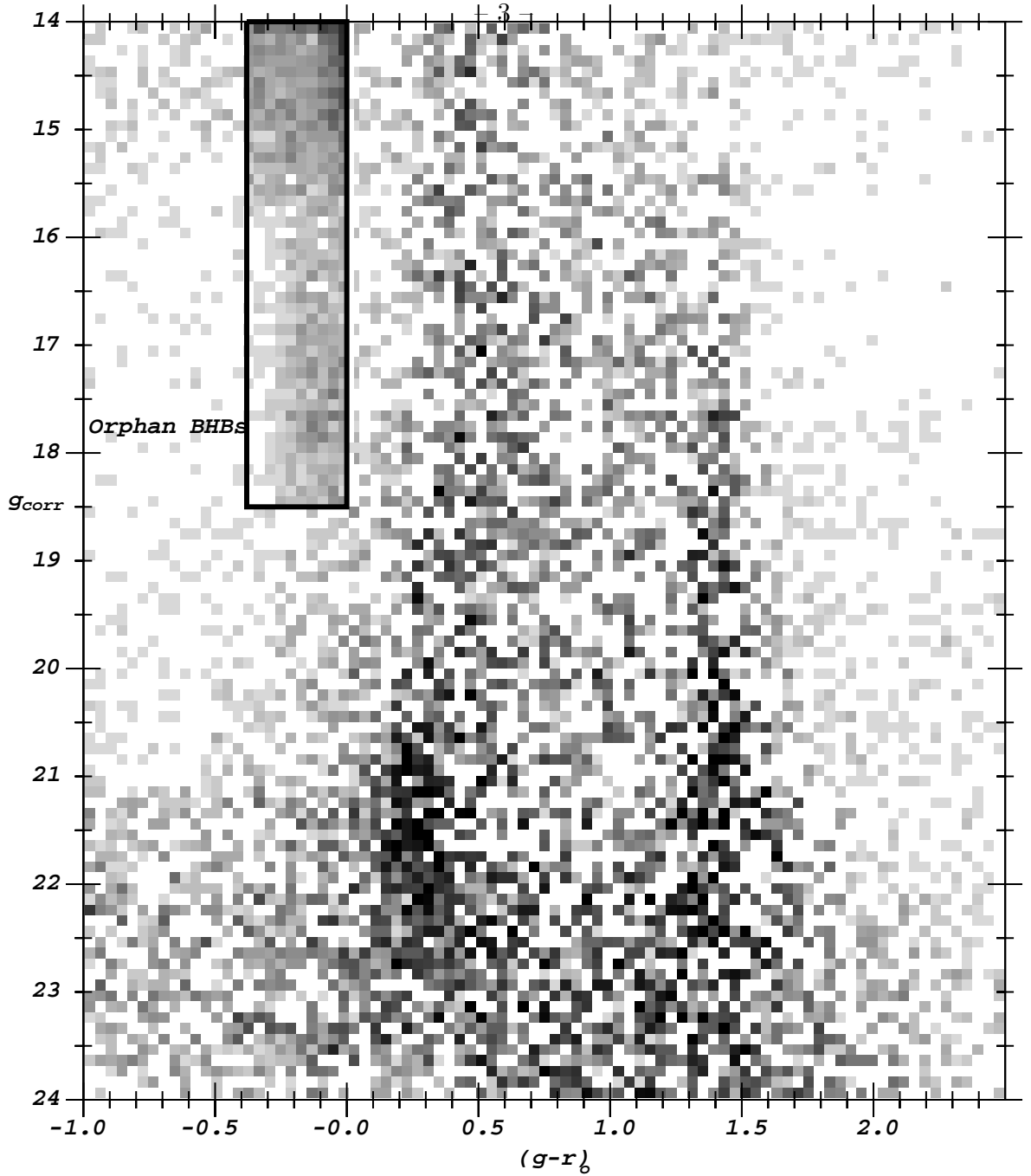}
\caption[CMD density diagram]{
\footnotesize
We present a color-magnitude Hess diagram of the stars within two degrees of the Orphan Stream
($-2^\circ<B_{\rm Orphan}<2^\circ$), and not near the Sagittarius dwarf tidal stream
($-26^\circ<\Lambda_{\rm Orphan}<-7^\circ$ or $10^\circ<\Lambda_{\rm Orphan}<40^\circ$).
A background population with the same $\Lambda_{\rm Orphan}$ range, but with
$2^\circ<|B_{\rm Orphan}|<4^\circ$, has been subtracted.
The apparent magnitudes of the stars have been shifted, based on their position along
the stream, so that BHBs at the distance of the Orphan Stream will have an apparent
magnitude of $g_{\rm corr}=17.75$.  This is the apparent magnitude of BHB stars at
$\Lambda_{\rm Orphan}=0^\circ$.  We identify the turnoff of the Orphan Stream in this
diagram at $(g_0, (g-r)_0) = (21.2, 0.22)$, at its bluest point.  
This figure includes an inset in which BHB 
stars have been color-selected to exclude QSOs and blue stragglers.  Note the excess of BHB stars
at $(g-r)_0=-0.15$ and $g_{\rm corr}=17.75$.  
Since there are very few color-selected BHB stars in the background
region, background is not subtracted in the inset.  
}
\end{figure}

\begin{figure} 
\includegraphics[scale=0.75,viewport=0in 4in 6in 14in]{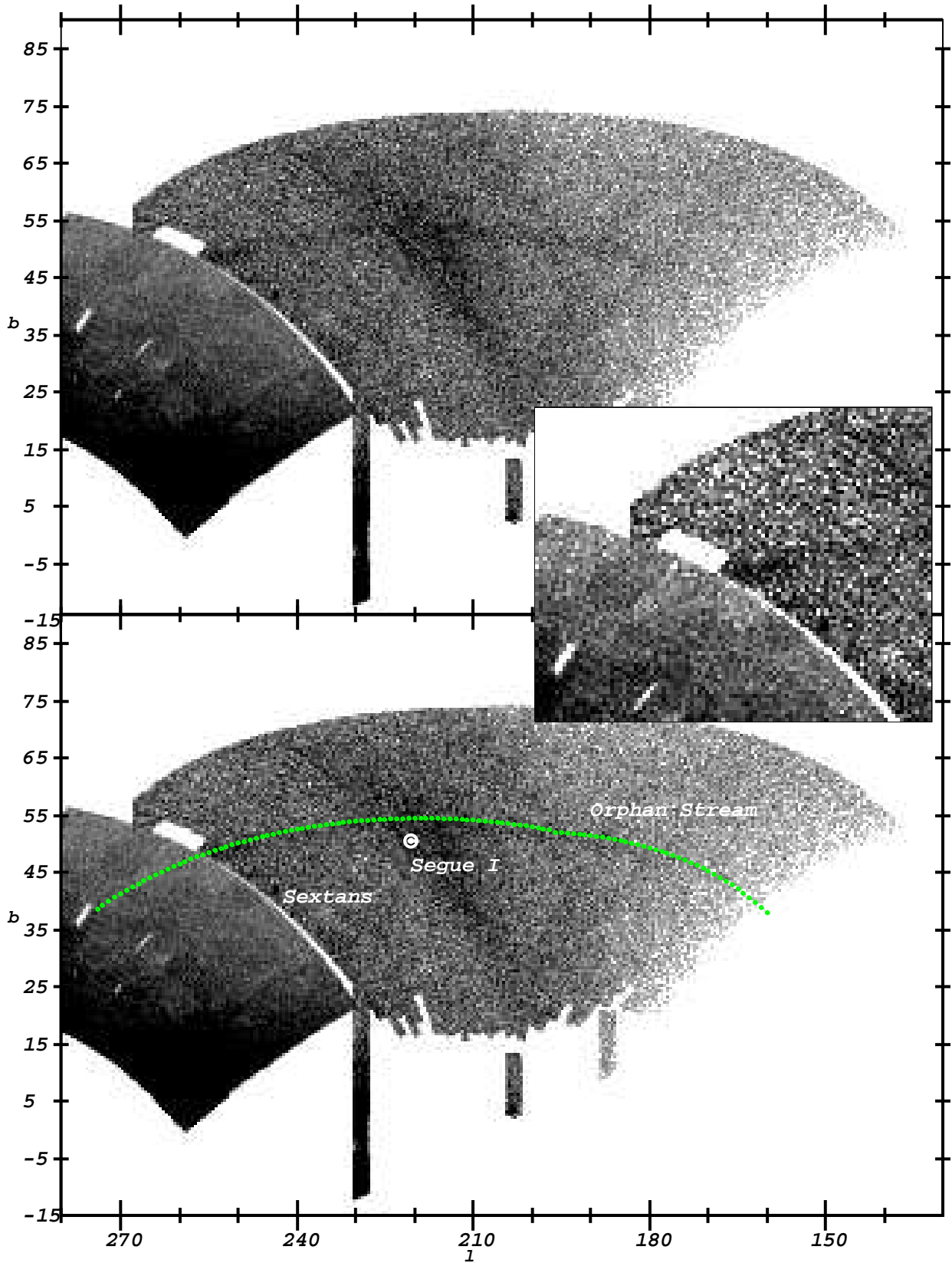}
\caption[Hess Diagrams of Orphan Stream]{
\footnotesize
Similar to Figure 1, except that the $g_0$ of stars have been shifted as a function
of $\Lambda_{\rm Orphan}$ based on the fit in Figure 2.  Stars with $20.7 < g_{\rm corr} < 21.7$, the
magnitude centered on the Orphan Stream's turnoff, are plotted in a Hess diagram over the SDSS footprint.  The stream can be seen extending to nearly $l = 170^\circ$.  
Additionally, $B_J,R_2$ data from SuperCOSMOS is extracted and added to
the figure to cover the region of the Orphan Stream where no SDSS data exists. The stars here 
have $20 < B_J < 21, 0.5 < B_J-R_2 < 1.0$ (with the default calibration and reddening corrections 
adopted throughout).  A narrow
trail is visible in the SuperCOSMOS extension, running from $(l,b) = (268^\circ,38^\circ)$ to $(255^\circ,48^\circ)$.
The lower panel adds the $B_{\rm corr}=0$ trace of the stream, which differs slightly from $B_{\rm Orphan}=0^\circ$ for $l < 200^\circ$.  The location of the
halo object Segue-1 is indicated.  The inset shows a blowup of the
region with $230^\circ < l < 280^\circ, 30^\circ < b < 70^\circ$.
}
\end{figure}

\begin{figure} 
\includegraphics[scale=0.65,angle=-90]{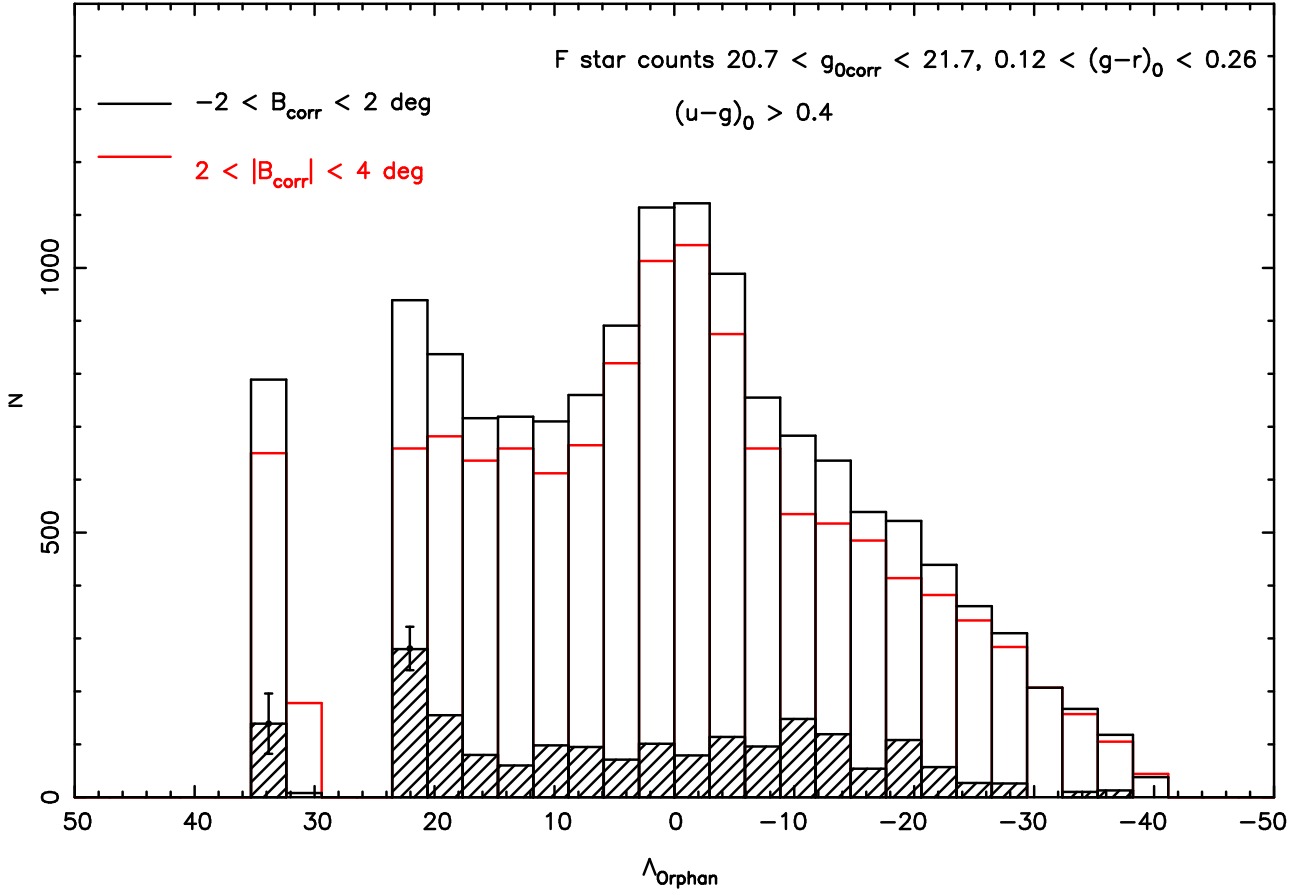}
\caption[Histogram of F star counts on and off stream]{
\footnotesize
Number counts of F turnoff stars within $\pm 2^\circ$ of the stream are plotted as an open black histogram.
The number of background turnoff stars off-the-stream on either side ($2^\circ < |B_{\rm corr}| < 4^\circ$) are plotted in red.  The difference is plotted as a hashed histogram.  Note the
significant excess of turnoff stars over background near $\Lambda_{\rm Orphan} = +23^\circ$, 
corresponding to $(l,b)=(255^\circ,49^\circ)$.  It is possible that 
the stream progenitor lies in this region of sky.  The bins at $\Lambda_{\rm Orphan} = 21^\circ, 33^\circ, \rm ~and~ 36^\circ$ have been corrected for incompleteness in both the data and background counts.
}
\end{figure}

\begin{figure} 
\includegraphics[scale=0.75,angle=-90]{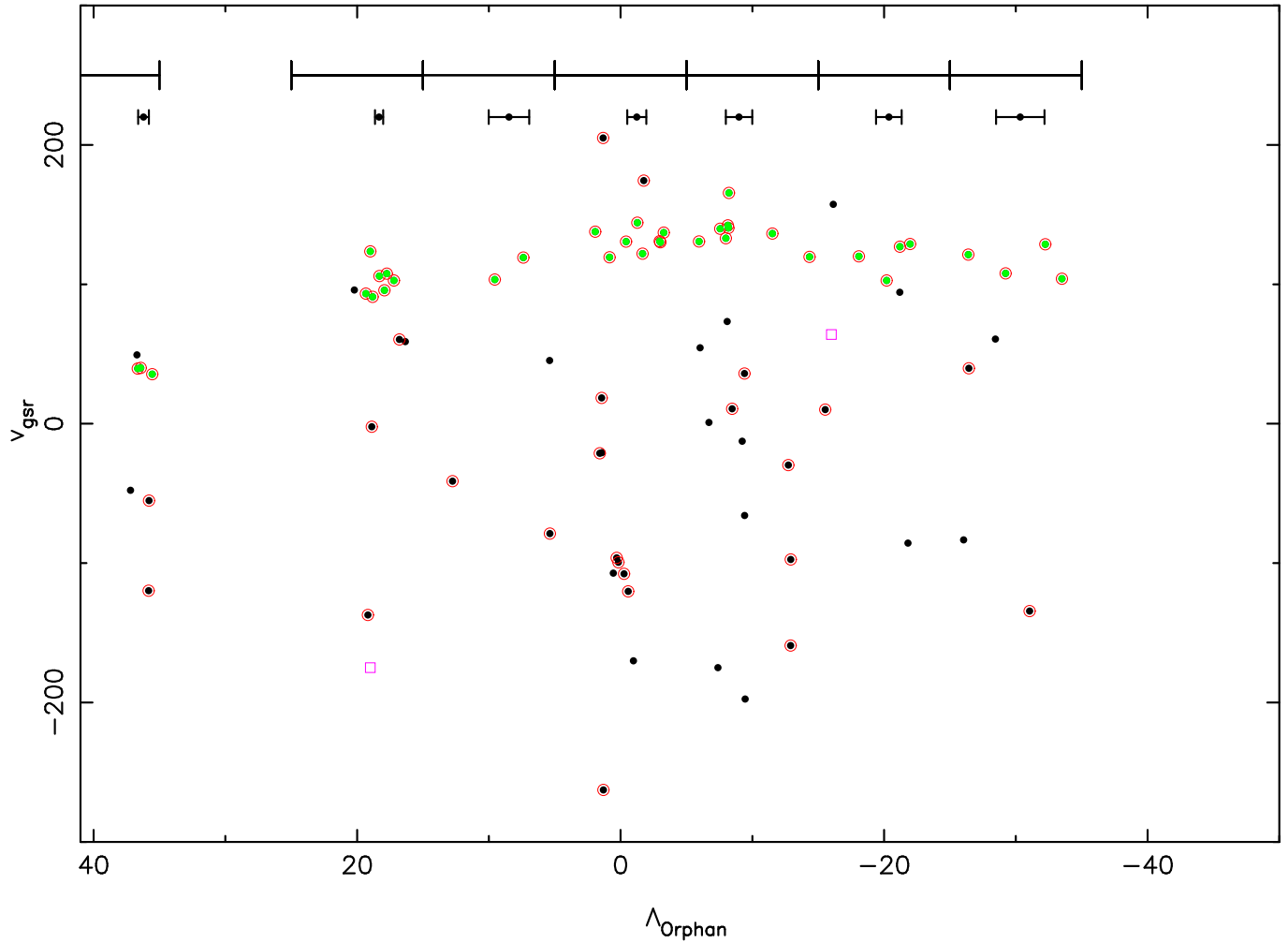}
\caption[Velocities of BHB stars within $2^\circ$ of the Orphan Stream.]{
\footnotesize
Velocities of BHB stars within $2^\circ$ of the Orphan Stream. The solid dots indicate
all stars with the correct surface gravity, $ugr$ colors and $g_0$ magnitudes 
(based on Figure 2) to be BHBs and stream members.  The red circles indicate 
stars with [Fe/H]$_{\rm WBG}$ that is lower than -1.6.
Points with green centers are our Orphan Stream candidate BHBs with spectra which
meet our final velocity cut as well as the metallicity cut.  The concentration of points 
at $\Lambda_{\rm Orphan} =0^\circ, v_{\rm gsr} \sim -100~\rm ~km~s^{-1}$ are likely 
associated with the Sagittarius leading tidal stream \citep{ynetal09b}. The 10$^\circ$ 
marks indicate the bins used in calculating the Orphan Stream properties at positions given
by the lower black points with error bars.
The magenta boxes indicate the earlier, preliminary velocity possibilities for the Orphan Stream
from \citet{betal07}, converted from radial velocity to $v_{\rm gsr}$.

}
\end{figure}

\begin{figure} 
\includegraphics[scale=0.65,angle=-90]{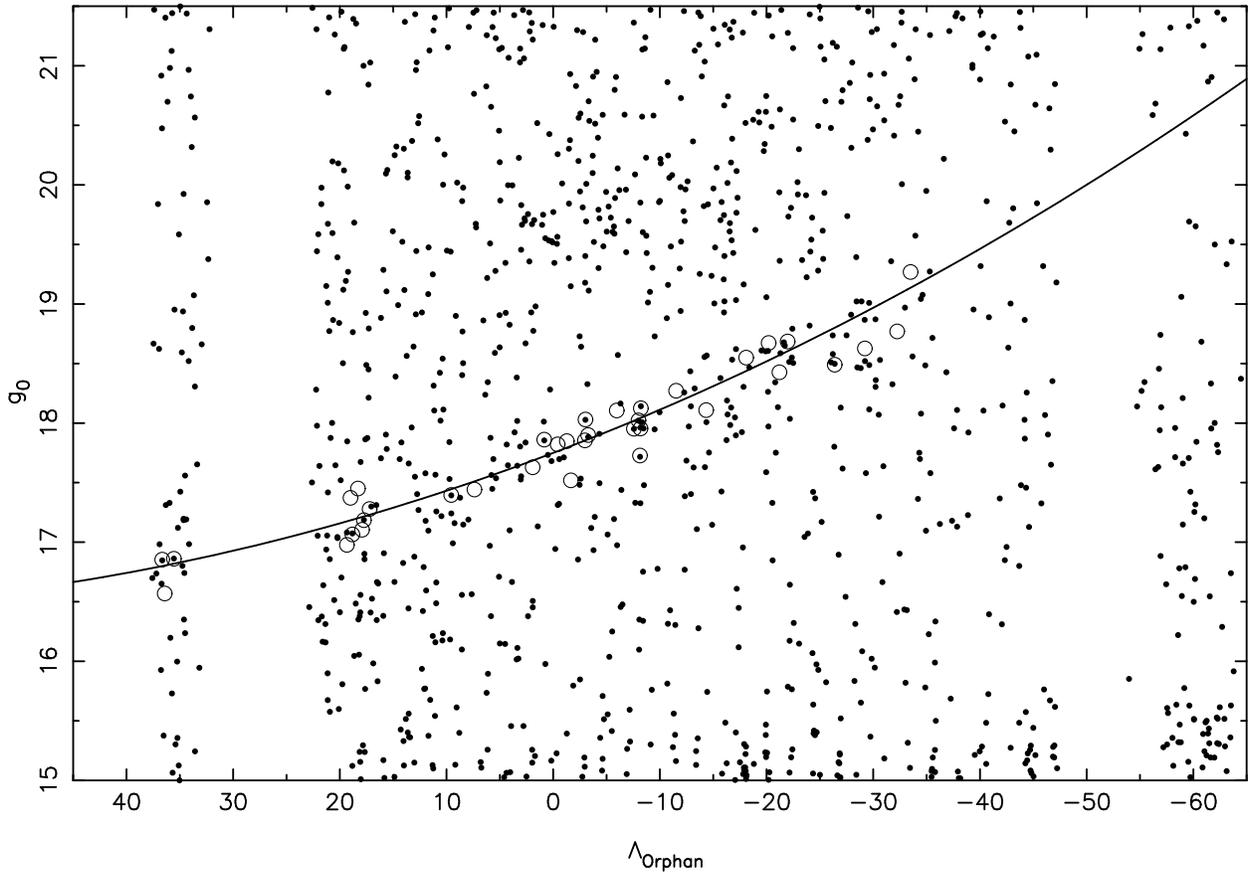}
\caption[Lambda vs. g]{
\footnotesize
Similar to Figure 2 except: 1) stars are selected to be within $2^\circ$ of the stream instead of just $1^\circ$ and 2) stars are chosen using the corrected $B_{\rm corr}$ for
$\Lambda_{\rm Orphan} < -15^\circ$.  Orphan Stream BHBs with spectra (green filled circles from 
Figure 6) are overlaid as open circles.  We also overlay the adopted fit to the $g_0$ magnitude
of the BHB stars as a function of $\Lambda_{\rm Orphan}$.
}
\end{figure}

\begin{figure}
\includegraphics[scale=0.75,viewport=0in 0in 6in 10in]{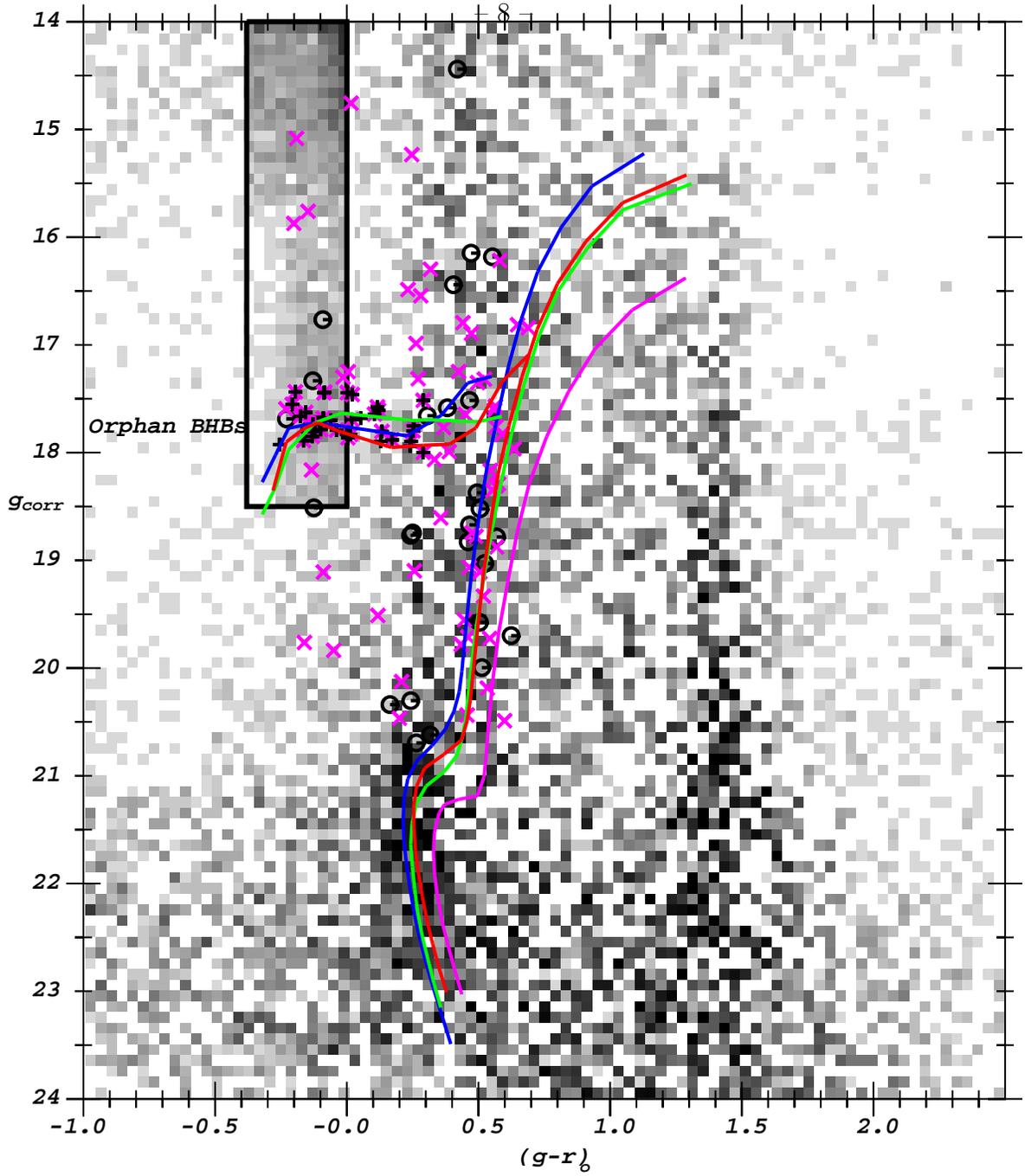}
\caption[CMD density diagram]{
\footnotesize
Same as Figure 3, with overlay of globular cluster fiducial sequences, normalized to place
their HBs at $g_{\rm corr} = 17.75$.  The clusters
are: M92 $\rm [Fe/H] = -2.3$ (blue), M3 $\rm [Fe/H] = -1.57$ (red), M13 $\rm [Fe/H] = -1.54$ 
(green) and M71 $\rm [Fe/H] = -0.73$ (magenta).  Orphan Stream BHB spectral candidates (filled green points
from Figure 6) are marked with small black $+$ signs.  Other spectra within $2^\circ$ of the
stream, and having velocity, metallicity and surface gravity consistent with the stream
are indicated with magenta crosses $\times$ .  Control field spectra, meeting the same 
velocity, gravity and metallicity conditions but which lie more than $2^\circ$ 
and less then $5^\circ$ from the stream are indicated with black circles.
}
\end{figure}

\begin{figure}
\includegraphics[scale=0.75,angle=-90]{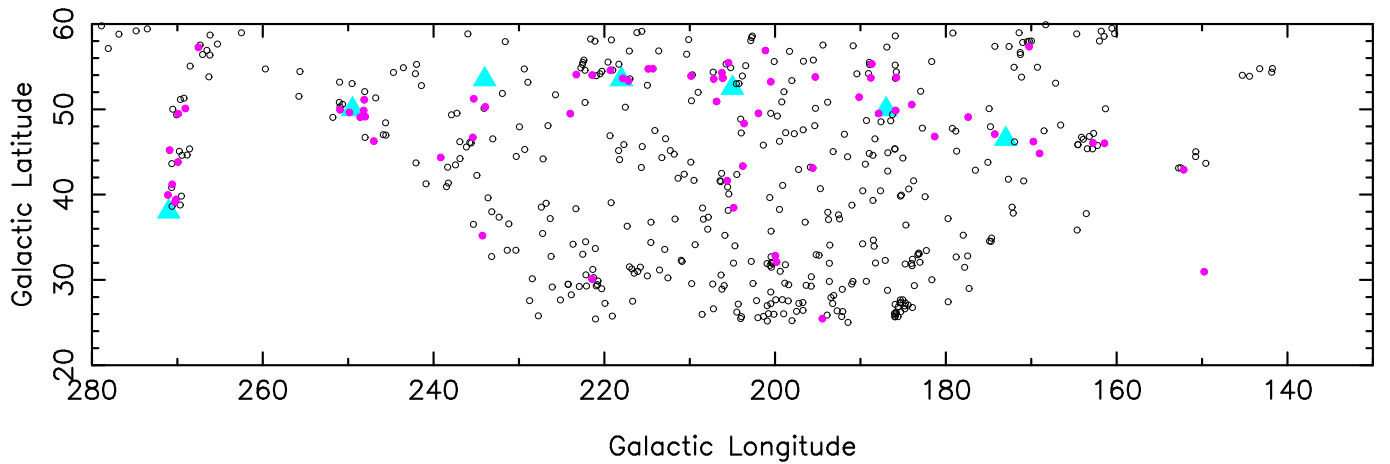}
\caption[Sky positions of Orphan Stream from BHB stars]{
\footnotesize
We show the sky positions for all Orphan Stream spectral BHB candidates which
meet the luminosity, velocity, color and magnitude selection conditions of Figures 
6 and 8 as open circles.  Filled magenta dots show those BHBs with exceptionally low 
metallicity, $\rm [Fe/H] < -1.6$. The blue triangles indicate the set of chosen 
fiducial positions along the Orphan Stream with spectral velocity measurements.  
There is good consistency between the low metallicity stars and the 
kinematically selected sample, except at $l = 270^\circ$.  
The SDSS data in this stripe are ambiguous, but we note that our extrapolated stream position
is a reasonable fit to the SuperCOSMOS data.
}
\end{figure}

\clearpage

\begin{figure} 
\includegraphics[scale=0.68,angle=-90]{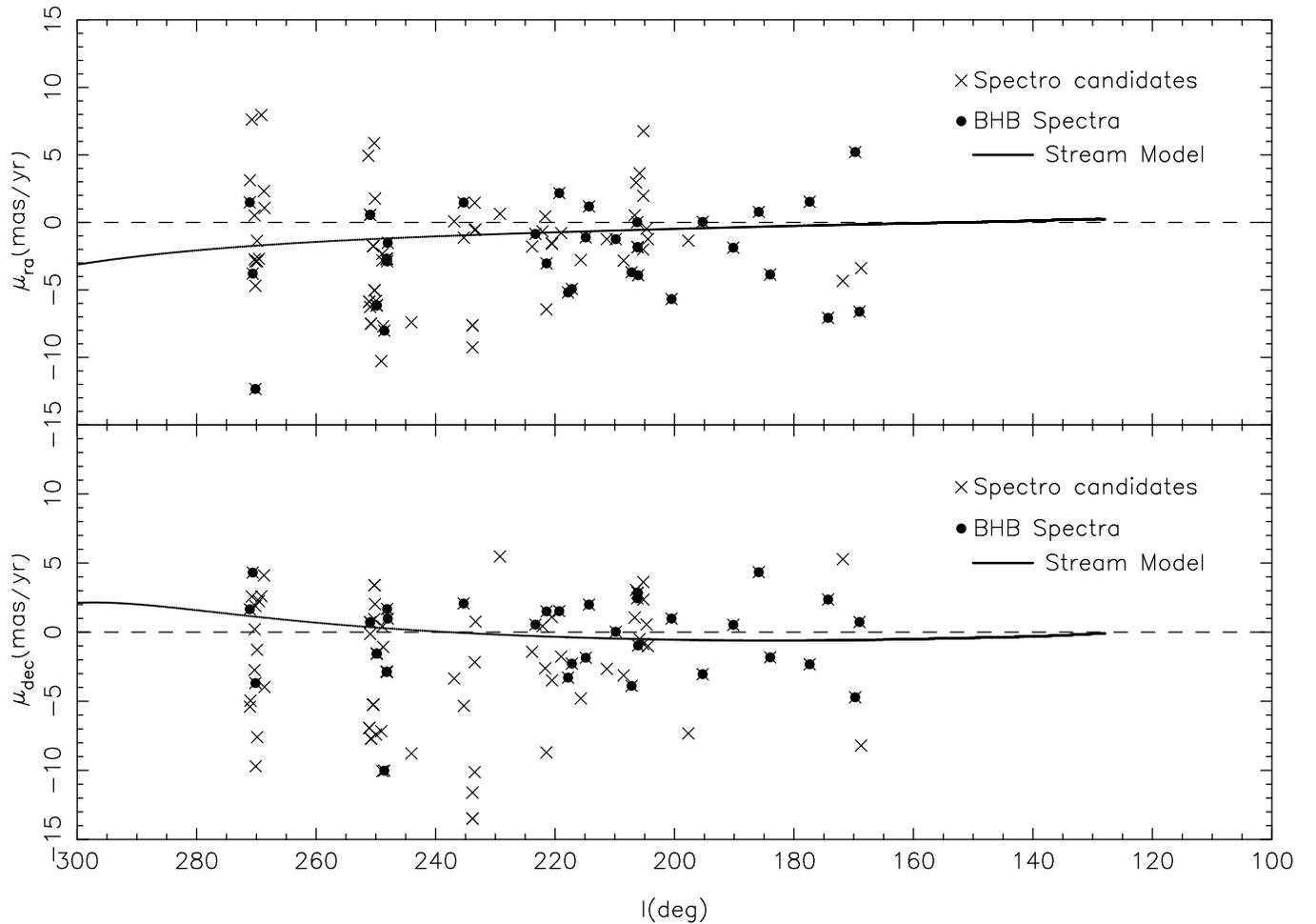}
\caption[Proper motions of Orphan Spec candidates]{
\footnotesize
Though proper motions of stream candidate stars are not used in fitting the
Orphan Stream orbit in this paper, they can provide a consistency check. 
We plot here $\mu_{\rm ra}$ and $\mu_{\rm dec}$ vs  $l$ for two subsets 
of candidate Orphan Stream stars:  
1) Spectroscopic stream candidates (magenta crosses from Figure 8), are shown as 
crosses, and 2) Spectroscopically more secure BHB stream candidates (green filled 
circles from Figure 6, black $+$ signs from Figure 8), are shown as filled black 
circles superimposed on a cross.  Errors on individual points are typically 
$3 ~\rm mas~yr^{-1}$ in both $\alpha$ and $\delta$.  Zero proper motion is indicated 
with a dashed line.  
A representative Orphan Stream orbit model (model 5) is overlaid as a heavy black line.
}
\end{figure}

\begin{figure} 
\includegraphics[scale=0.75]{fig11.ps}
\caption[Metallicities of Spectroscopic Orphan Candidates]{
\footnotesize
In the upper panel, we show the metallicity distribution of Orphan Stream BHB candidates, 
without any prior metallicity cut, as measured by the \citet{wbg99} method of estimating [Fe/H] based on Ca II K line
strength and $(g-r)_0$ color.  The Orphan Stream BHBs have a very low
metallicity of $\rm [Fe/H] = -2.1\pm 0.1$.
The lower panel shows the `adopted' SSPP [Fe/H], for the stars which meet the
magnitude, color, surface gravity and velocity cuts for stream membership (BHBs excluded).
The distribution for stars with $|B_{\rm corr}| < 2^\circ$ are shown in light outline,
while that for stars with $2^\circ < |B_{\rm corr}| < 4^\circ$ is shown in heavy
outline.
}
\end{figure}

\begin{figure} 
\includegraphics[scale=0.75]{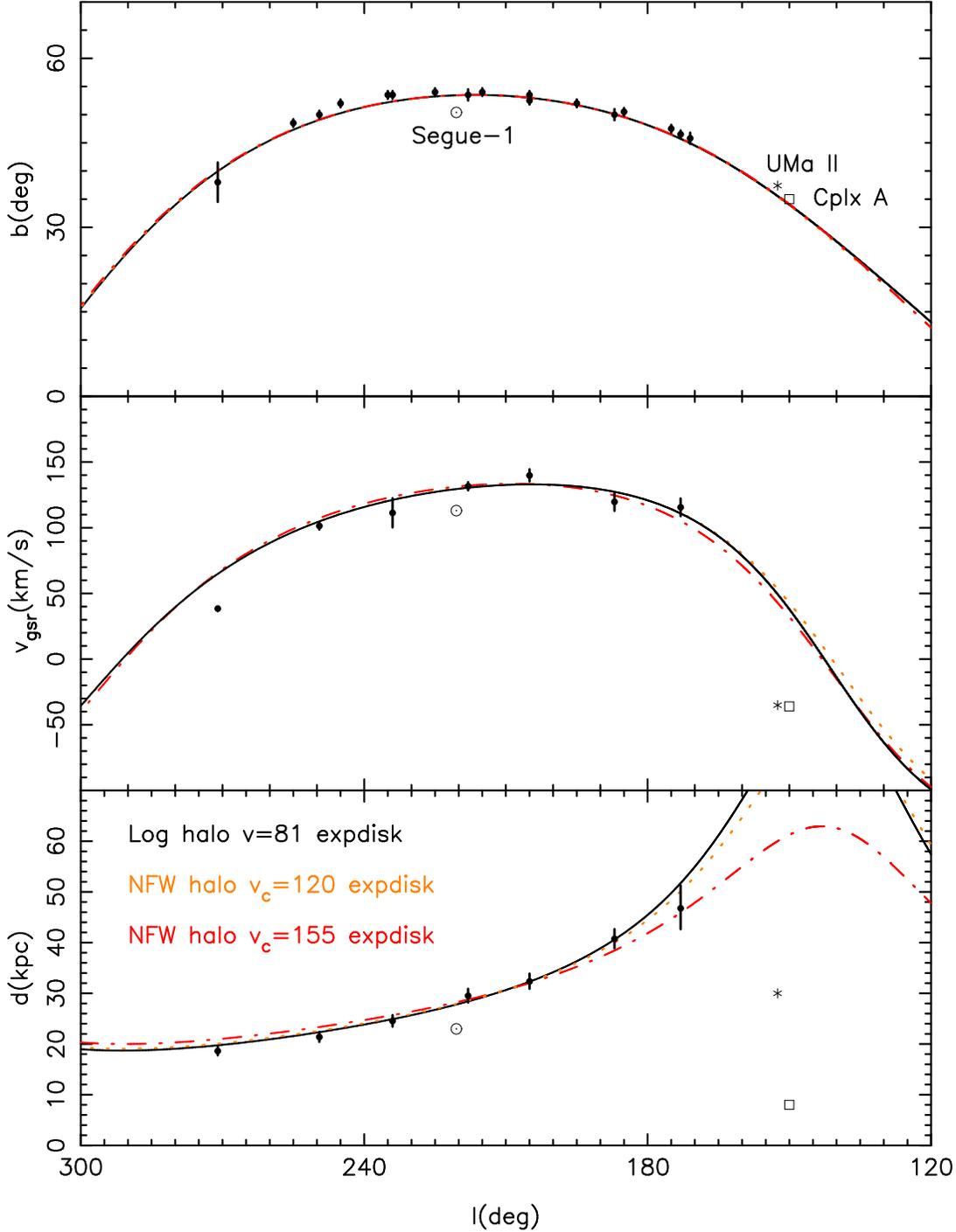}
\label{orbitfig}
\caption[Orbit]{
\footnotesize
We present orbit fits ($b, v_{gsr}, and d$ vs. Galactic longitude) to the Orphan Stream 
data for three different halo potentials (models 1-3), with a
fixed bulge and exponential disk model with parameters copied from \citet{xetal08}.
In red (dot dash, model 1) we show the best orbit using the NFW halo parameters from \citet{xetal08}.  In orange
(dotted, model 2) we show the best fit if we allow the $v_c$ parameter to vary in the NFW profile.  In black 
(solid, model 3) we show the best fit if we instead use a logarithmic potential.
Note that the best fit NFW and logarithmic potentials give similar fits, but both have a lower
velocity (and therefore halo mass) than found by \citet{xetal08}.
To fit the Orphan Stream distance data, it is necessary to reduce the amplitude of the halo 
potential by about 40\%. The distance-velocity space locations of Segue-1, Ursa Major II, Complex A 
are indicated as labeled in the top panel. Note that only Segue-1 is possibly associated with this 
tidal stream.
\footnotesize
}
\end{figure}

\begin{figure} 
\includegraphics[scale=0.75,angle=-90]{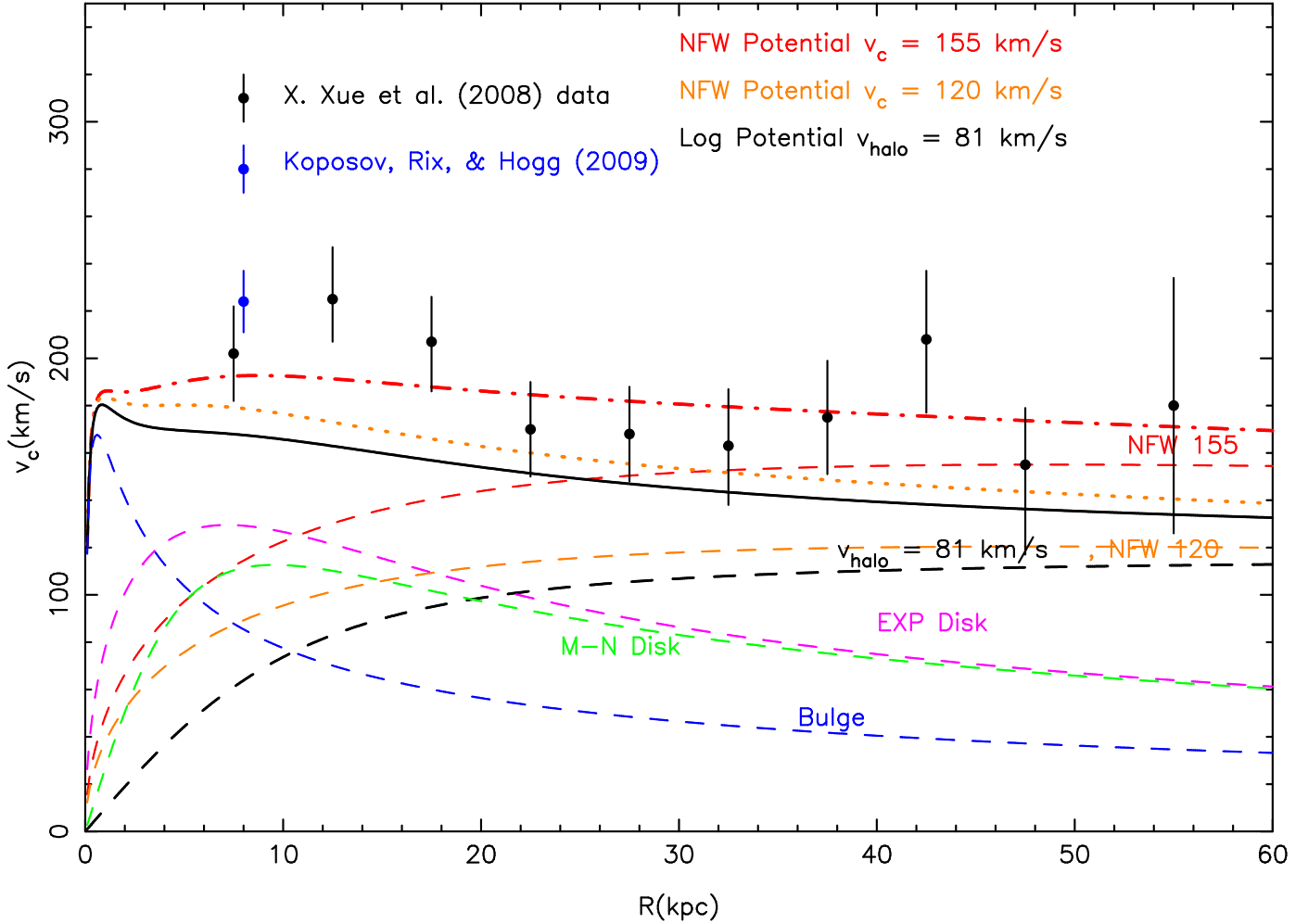}
\caption[Circular Velocity curves]{
\footnotesize
We now compare the best fit models from Figure 12 with the Milky Way rotation curve data from \citet{xetal08}
and \citet{krh09}.  The red curve (dot dash, model 1) is the best fit found by \citet{xetal08}, from
Figure 16a of that paper,
so it is a good fit to the data points.  The orange curve (dotted, model 2) and black curve (solid,
model 3), for which we allowed the halo mass to vary, are significantly below all of the rotation 
curve data points.  The rotation curves from individual components of the potential (exponential 
disk, bulge, and NFW halo) used in \citet{xetal08} are also shown. 
We also show a Miyamoto-Nagai disk, scaled to a total mass similar to
that of the exponential disk, to show that the rotation curve in the region we probe is not
dramatically different for different disk profiles, and the halo rotation curves for the lower
mass halos.
}
\end{figure}

\begin{figure} 
\includegraphics[scale=0.75]{fig14.ps}
\caption[Orbit]{
\footnotesize
We present orbit fits ($b, v_{gsr},$ and $d$ vs. Galactic longitude) to the Orphan Stream
data for potential models 4-7, with a
fixed bulge and \citet{1975PASJ...27..533M} disk with parameters copied from \citet{ljm05}.
In green (dash, model 4) we show the fit best orbit using the logarithmic halo parameters 
from \citet{ljm05}.  In red (dot dash, model 6) we show the best fit 
with NFW halo parameters fixed from \citet{xetal08}.
Neither of these fixed halo potentials from previous papers gives a good fit to the relative
distances along the Orphan Stream; the slope of the $d$ vs. $l$ plot is too shallow for
these models.  The black (solid, model 5) and orange curves (dotted, model 7) show the best fit
orbits if the disk is fixed and the halo masses are allowed to vary for the logarithmic
and NFW profile halos, respectively.  As in Figure 12, we see very little difference between
the best logarithmic and NFW profile fits, and the preferred values of the halo mass is
lower than that of previous authors.  Again, it is necessary to reduce the amplitude of the halo
potential by about 40\% to fit the Orphan Stream distances. The locations 
of Segue-1, Ursa Major II, Complex A are as labeled as in Figure 12.
}
\end{figure}

\begin{figure} 
\label{rotcurves}
\includegraphics[scale=0.75,angle=-90]{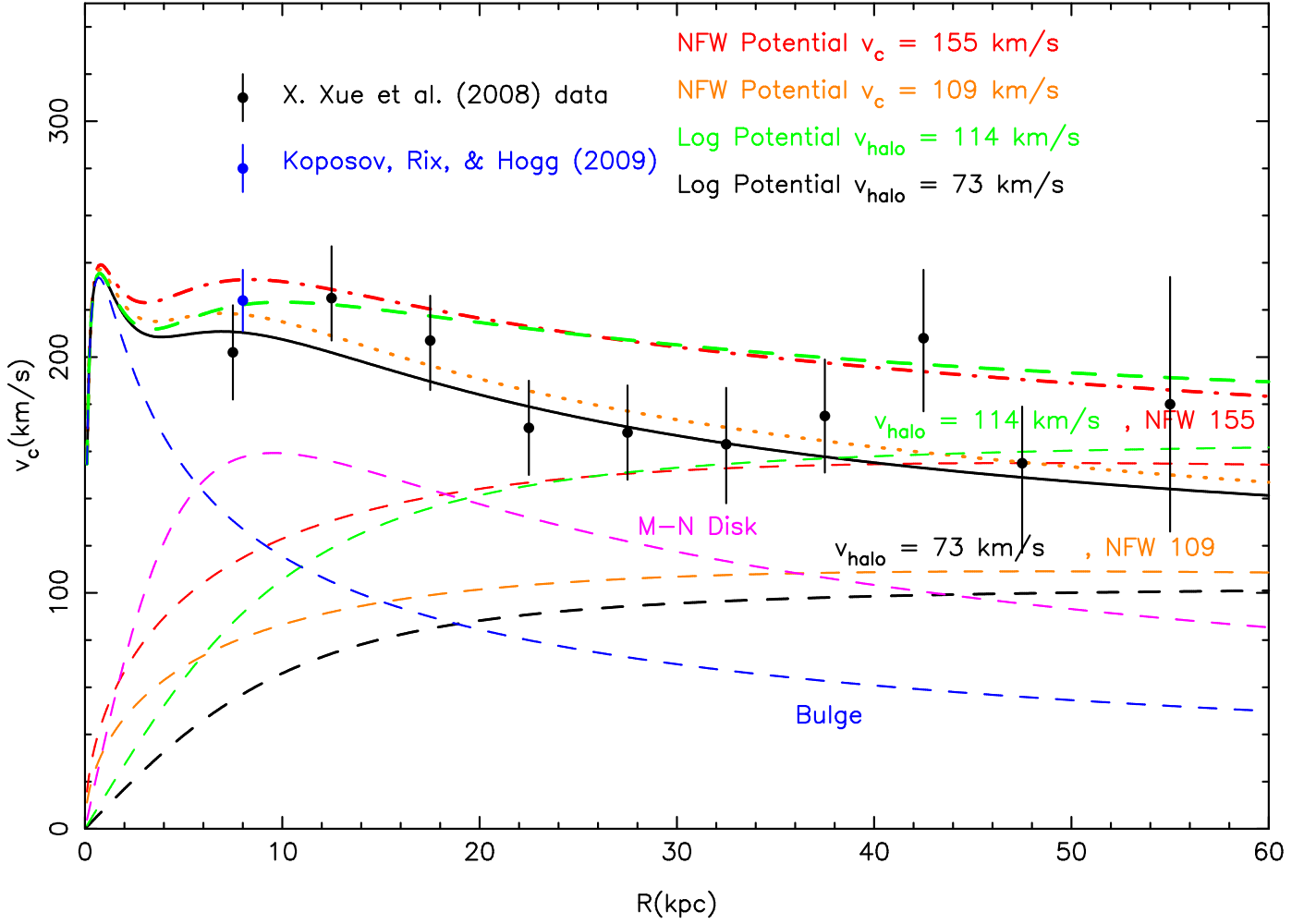}
\caption[Circular Velocity curves]{
\footnotesize
We now compare the best fit models from Figure 14 with the Milky Way rotation curve data from \citet{xetal08}
and \citet{krh09}.  All of the models use a fixed M-N disk and bulge, with a mass about twice as
large as the exponential disk in the models in Figures 12 and 13.  The green (dash, model 4) and
red (dot dash, model 6) curves use fixed parameters from the logarithmic halo potential of \citet{ljm05}
and NFW halo potential of \citet{xetal08}, respectively.  Neither of these fit the Orphan Stream
distances, and they are not especially good fits to the rotation curve, predicting rotation
velocities above most of the data points.  The black (solid, model 5) and orange curves (dotted, model
7) are reasonably good fits to the \citet{xetal08} data, even though they were not fit to this
data.  In these models, the halo mass was allowed to vary.  Our best fit model is model 5 (black, solid
curve), though model 7 is nearly as good.
The rotation curves from individual components of the potential (M-N
disk, bulge, and NFW or logarithmic halo) are also shown. 
}
\end{figure}

\begin{figure} 
\includegraphics[scale=0.75]{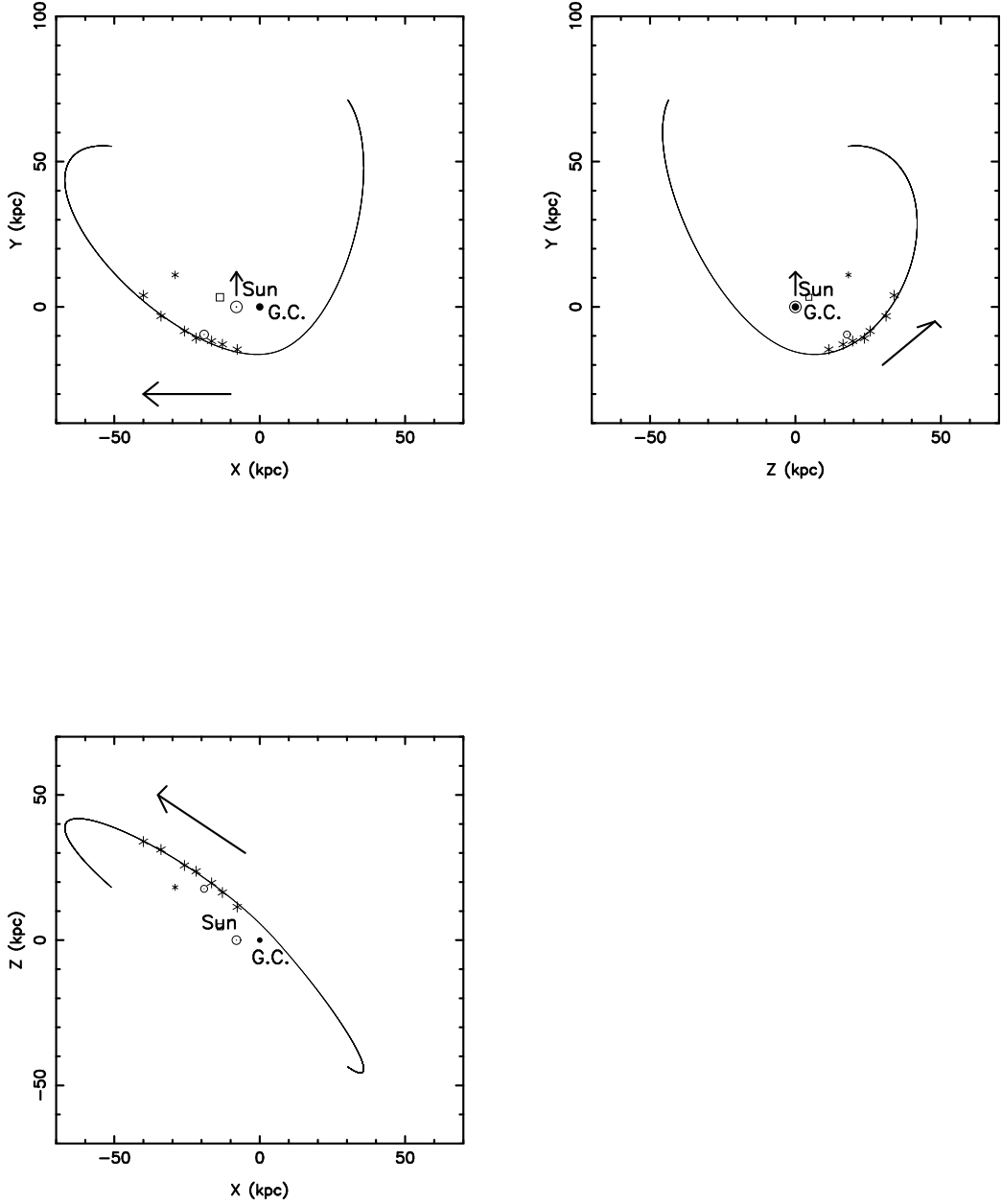}
\caption[Orbit]{
\footnotesize
The black line shows our preferred fit orbit in the logarithmic halo with 
$v_{\rm halo} = 73 \rm ~km \rm ~s^{-1}$ in right-handed Galactic rectangular coordinates 
$(X,Y,Z)$. The Sun is at $(-8,0,0) \rm ~kpc$ and the Galactic center at the origin. The 
$(l,b)$ coordinates and BHB magnitudes are converted to $(X,Y,Z)$ assuming a BHB absolute 
magnitude of $M_g = 0.45$. The arrows indicate the forward direction of the orbit and the 
Sun's motion.  Points from Table 2 of observations of the Orphan Stream data are shown 
as asterisks.  The positions of Segue-1, UMa II and Complex A are shown with the same 
symbols as in Figure 14.
}
\end{figure}

\begin{figure}
\label{orbitnbody} 
\includegraphics[scale=0.75]{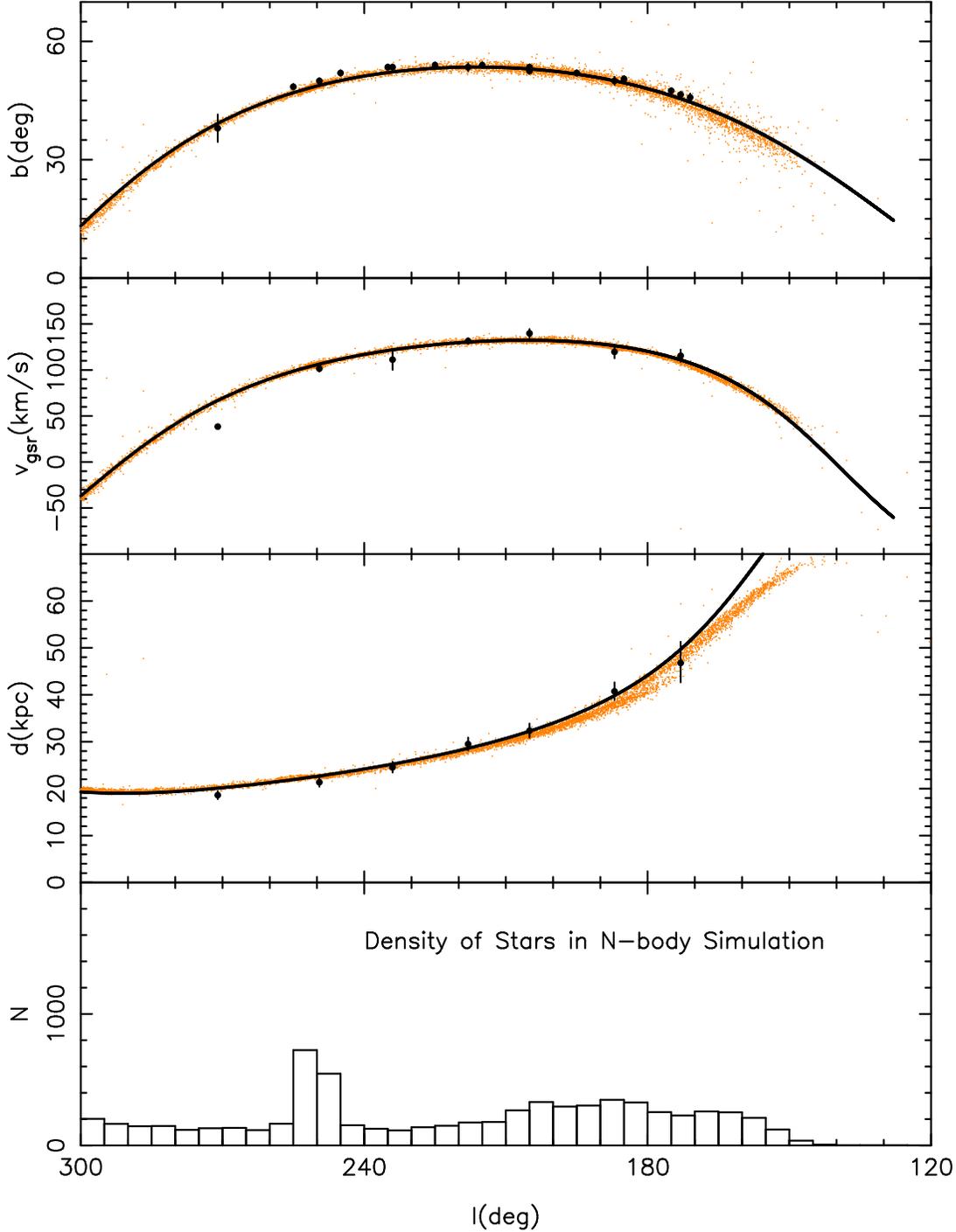}
\caption[Orbit]{
\footnotesize
The $b$, $v_{\rm gsr}$ and $d_{\rm Sun}$ vs. $l$ (top 3 panels) orbit of the 
preferred halo (logarithmic potential with $v_{\rm halo} = 73~\rm km~s^{-1}$, M-N disk and heavy bulge), simply integrated, is shown as a heavy 
black curve. A 10,000 point 
N-body simulation of a Plummer sphere dwarf with $M_{\rm total} \sim 10^6 M_\odot$ 
is integrated in this potential from a time 4 Gyr ago forward 
for 3.945 Gyr placing the proposed progenitor at $(l,b) \sim (250^\circ,50^\circ)$.  
The lower panel shows the density distribution of the final (current) epoch positions
of the N-body points as a histogram in $l$.  Note the similarity of the 
the simulated distribution, including the dip in density 
at $l=240^\circ (\Lambda_{\rm Orphan} = +15^\circ$) seen in Figure 5, and the spread (first panel) 
and density falloff at $l < 180^\circ$ seen in Figure 4. 
}
\end{figure}

\end{document}